\documentclass[numberedappendix]{emulateapj}

\newcommand{\ba}{\begin{eqnarray}}
\newcommand{\ea}{\end{eqnarray}}
\newcommand{\beq}{\begin{equation}}
\newcommand{\eeq}{\end{equation}}
\newcommand{\g}{\gamma}
\newcommand{\gp}{{\gamma^\prime}}

\newcommand{\dD}{{\delta_{\rm D}}}

\def\e{\epsilon}
\def\ep{\epsilon^\prime}

\def\asympt{\buildrel \rightarrow \over {_{u\gg 1}}}

\usepackage{natbib}

\begin{document}

\title{Equipartition Gamma-Ray Blazars and the Location of the Gamma-Ray Emission Site in 3C 279}
%\footnote{author list in alphabetical order}

\author{Charles D. Dermer\altaffilmark{1}, Matteo Cerruti\altaffilmark{2,3}, Benoit Lott\altaffilmark{4}, Catherine Boisson\altaffilmark{3}, Andreas Zech\altaffilmark{3} }

\altaffiltext{1}{Code 7653, Space Science Division, U.S. Naval Research Laboratory, Washington, DC 20375, USA. e-mail: charles.dermer@nrl.navy.mil}
\altaffiltext{2}{Harvard-Smithsonian Center for Astrophysics; 60 Garden Street, 02138 Cambridge, MA, USA. email: matteo.cerruti@cfa.harvard.edu  }
\altaffiltext{3}{Laboratoire Univers et THeories (LUTH), Observatoire de Paris-Meudon, 5 Place Jules Janssen 92195 Meudon Cedex, France. }
\altaffiltext{4}{Centre d'\'Etudes Nucl\'eaires Bordeaux Gradignan, Universit\'e de Bordeaux, CNRS/IN2P3, UMR 5797, Gradignan, 33175, France}

\begin{abstract}
%Multiwavelength campaigns involving the  Large Area Telescope (LAT) on the {\it Fermi Gamma ray Space Telescope} are
%providing the most detailed blazar spectral energy distributions (SEDs) to date.
%Generally, radio galaxies and high-synchrotron-peaked BL Lac objects are well fit with 
%synchrotron self-Compton (SSC) models. In contrast, low synchrotron-peaked objects, 
%especially flat spectrum radio quasars (FSRQs), require the addition of an 
%external-Compton scattering component. 
Blazar spectral models generally have numerous unconstrained parameters, 
leading to ambiguous values for physical properties like Doppler factor $\dD$ or fluid magnetic field $B^\prime$. 
To help remedy this problem, a few modifications of the standard leptonic blazar jet scenario are considered. 
First, a log-parabola function for the electron distribution is used.  Second, 
analytic expressions relating energy loss and kinematics to blazar luminosity and variability, written 
in terms of equipartition parameters,  imply $\dD$, $B^\prime$, and the 
peak electron Lorentz factor $\gamma_{pk}^\prime$.
%, possibly compatible with a second-order Fermi acceleration scenario, is assumed. 
The external radiation field in a blazar is approximated by Ly$\alpha$ radiation from the broad line region (BLR) 
and $\approx 0.1$ eV infrared radiation from a dusty torus. 
When used to model  3C 279 SEDs from 2008 and 2009 reported by Hayashida et al. (2012), we derive $\dD \sim 20$ -- 30, $B^\prime \sim$ few G,
and total (IR + BLR) external radiation field energy densities $u\sim 10^{-2}$ -- $10^{-3}$ erg cm$^{-3}$,
implying an origin of the $\gamma$-ray emission site in 3C 279 at the outer edges of the BLR. This is
 consistent with the $\gamma$-ray emission site being located 
at a distance $R \lesssim \Gamma^2 c t_{var} \sim 0.1 (\Gamma/30)^2 (t_{var}/10^4{\rm~s})$ pc from the 
black hole powering 3C 279's jets, where $t_{var}$ is the variability time scale of the radiation in the source frame, 
and at farther distances for narrow-jet and magnetic-reconnection models. 
Excess $\gtrsim 5$ GeV $\gamma$-ray emission observed with Fermi LAT from 3C 279 challenge
the model, opening the possibility of non-leptonic and hadronic origin of the emission.
For low hadronic content, 
absolute jet powers of $\approx 10$\%
of the Eddington luminosity are calculated.
\end{abstract}


%%%%%%%%%%%%%%%%%%%%%%%%%%%%%%%%%%%%%%%%%%%%%%%%%%%%%%%%%%%%%%%%%%%%%%
%%%%%%%%%%%%%%%%%%%%%%%%%%%%%%%%%%%%%%%%%%%%%%%%%%%%%%%%%%%%%%%%%%%%%%
\section{Introduction} \label{sec:introduction}
%%%%%%%%%%%%%%%%%%%%%%%%%%%%%%%%%%%%%%%%%%%%%%%%%%%%%%%%%%%%%%%%%%%%%%
%%%%%%%%%%%%%%%%%%%%%%%%%%%%%%%%%%%%%%%%%%%%%%%%%%%%%%%%%%%%%%%%%%%%%%

The time-varying emission spectrum of a blazar encodes important
information about particle acceleration and radiation, jet structure
and environment, and the mechanisms by which black holes power jets.
Determining the jet physics and environment from the
spectral energy distributions (SEDs) of blazars 
is a tricky problem of inversion. 
Part of the problem is that observational limitations
make it difficult to obtain a detailed picture of the blazar SED, as 
the blazar continuum extends over some 20 orders of magnitude from MHz radio frequencies to TeV $\gamma$-ray 
energies, and therefore requires coordinated campaigns at many 
wavelengths. The observing program furthermore has to be as nearly simultaneous as possible, 
due to the highly variable nature of blazars. 
In spite of these obstacles, extraordinary progress has been made to 
obtain a relatively complete description of the blazar SED.

Here we introduce a new model of blazar SEDs, built on the principle of 
equipartition. We use the 3C 279 data in  \citet{2012ApJ...754..114H} from which
to make a test of the model. But first we give background. 
 
The EGRET instrument on the {\it Compton} Gamma Ray Observatory first showed 
that 100 MeV -- GeV  emission often dominates the apparent power output of blazars during flaring 
states of flat spectrum radio quasars (FSRQs), and that the $\gamma$ rays
form a spectral component that is distinct from the nonthermal synchrotron emission
radiated at radio, IR and optical frequencies \citep[see, e.g.,][]{1992ApJ...385L...1H,1998ApJ...497..178W}. The ground-based
Cherenkov $\gamma$-ray telescopes demonstrated that the BL Lac class 
are powerful sources of TeV emission \citep{1992Natur.358..477P}, and that FSRQs, including 3C 279 \citep{2008Sci...320.1752M}
at redshift $z = 0.5362$  \citep{1996ApJS..104...37M}, 
PKS 1510-089 ($z = 0.36$) \citep{2010HEAD...11.0705W,2012ATel.3965....1C,2013A&A...554A.107H}, and 4C +21.35 (PKS 1222+216; z = 0.432) \citep{2011ApJ...730L...8A}, are 
sources of very-high energy (VHE; $\gtrsim 100$ GeV) $\gamma$ rays.

The Large Area Telescope (LAT) on  {\it Fermi}, 
which nominally operates in scanning mode, has revolutionized our understanding
of blazars by providing uninterrupted light curves and detailed SEDs of blazars. In campaigns
involving radio, IR, optical, X-ray, and ground-based $\gamma$-ray telescopes, 
superbly detailed SEDs have been measured that challenge our theoretical understanding. 
A few examples  include the high-synchrotron-peaked BL Lacs Mrk 421 \citep{2011ApJ...736..131A} and 
Mrk 501 \citep{2011ApJ...727..129A}, the intermediate synchrotron-peaked BL Lac 3C 66A \citep{2011ApJ...726...43A}, and 
the FSRQs 3C 454.3 \citep{2009ApJ...699..817A} and 3C 279 \citep{2010Natur.463..919A}. Mrk 421 and Mrk 501 are reasonably well fit with a leptonic synchrotron/SSC model, whereas 
an additional $\gamma$-ray component, thought to arise from Compton scattering
of external radiation fields,  is required in FSRQs and 
low- and intermediate synchrotron-peaked BL Lac objects.
 
A standard blazar paradigm consisting of magnetized plasma ejected with relativistic speeds in a collimated outflow along the polar axes of a rotating black hole has been developed to explain the overall observational features of blazars \citep[see][for a recent detailed review]{2012rjag.book.....B}. But even the simple synchrotron/SSC model suffers from a proliferation of parameters, which greatly increases with the introduction of external radiation fields having an uncertain spectrum and energy density. The range of possible fits to the SEDs makes it difficult to extract unique and therefore meaningful blazar jet properties. Part of this difficulty has to do with characterizing the nonthermal electron spectrum, and another part with the range of possibilities for the sources of target photons, which can include the accretion-disk \citep{1992A&A...256L..27D,1993ApJ...416..458D,2002ApJ...575..667D}, broad line region \citep[BLR;][]{1994ApJ...421..153S}, and molecular torus  \citep{2000ApJ...545..107B,2002A&A...386..415A,2005ApJ...629...52S}  \citep[see also][]{2007Ap&SS.309...95B,2009ApJ...704...38S}. 

To help alleviate these problems, we adopt a specific model of the electron 
distribution and assume that the primary target radiation
fields are (i) the BLR atomic-line radiation, and (ii) the IR radiation from a dusty torus
surrounding the nucleus. We employ 
a log-parabola function in Section 2 to describe the electron energy distribution in the
comoving fluid frame. Equipartition relations between the energy densities of the 
magnetic field, the nonthermal electrons, and the observed properties of the blazar synchrotron 
spectrum are
derived in Section 3. In Section 4, model results are shown for target photons of the external radiation  described 
 by monochromatic frequencies and various energy densities.
In Section 5, spectral fitting of recent SEDs of 3C 279 \citep{2012ApJ...754..114H} are used
to illustrate the method. Discussion of results is found in Section 6.

In a companion paper \citep{cer13}, we improve the 
model for the external radiation field, replacing  monochromatic infrared emission
 by thermal spectra, and the BLR radiation field by
a complex of atomic emission lines rather than by the Ly$\alpha$ line alone. 
We show that scattering of the BLR radiation 
under equipartition conditions can explain the GeV spectral cutoff discovered with 
{\it Fermi}-LAT in 3C 454.3 and other FSRQs and low- and intermediate-synchrotron-peaked 
BL Lac  objects \citep{2009ApJ...699..817A,2010ApJ...710.1271A}.

%%%%%%%%%%%%%%%%%%%%%%%%%%%%%%%%%%%%%%%%%%%%%%%%%%%%%%%%%%%%%%%%%%%%%%
%%%%%%%%%%%%%%%%%%%%%%%%%%%%%%%%%%%%%%%%%%%%%%%%%%%%%%%%%%%%%%%%%%%%%%
\section{Log-Parabola Electron Distribution} \label{sec:logpar}
%%%%%%%%%%%%%%%%%%%%%%%%%%%%%%%%%%%%%%%%%%%%%%%%%%%%%%%%%%%%%%%%%%%%%%
%%%%%%%%%%%%%%%%%%%%%%%%%%%%%%%%%%%%%%%%%%%%%%%%%%%%%%%%%%%%%%%%%%%%%%

In leptonic blazar jet models, the synchrotron and Compton $\gamma$-ray 
emission components  are radiated by nonthermal electron distributions that
are assumed to be isotropic in the jet fluid frame.
One technique is to fit the data by injecting power-law particle distributions
and allowing the particles to evolve in response to radiative and adiabatic energy losses
\citep[e..g.,][]{2002ApJ...581..127B,2003A&A...406..855M,ksk03}.
Many parameters must be specified, 
including the cutoff energies, injection indices,
and power, but this method is potentially useful 
to follow the dynamic spectral behavior of blazars.

Contrary to this approach, we 
 abandon any preconceptions about particle acceleration, and 
employ the simplest functional form that is able to provide reasonably good fits
to the snapshot SED data \citep[compare][ for SSC modeling 
of TeV blazars with power-law electron distributions.]{2008ApJ...686..181F}. 
%Only when we are unable  to obtain good fits to the data will we be forced to 
%use more complicated functions. 
For this purpose, we assume 
that the log-parabola function  
\begin{equation}
\log[\g^{\prime 2} N_e^\prime(\gp)]=\log[\gamma_{pk}^{\prime 2}N_e(\g^\prime_{pk})]-b\log(\g^\prime/\g^\prime_{pk})^2
\label{g2Ng}
\end{equation}
gives a good description of the
nonthermal electron Lorentz-factor ($\gamma^\prime$) distribution $N^\prime_e(\gp)$
in the comoving fluid frame, where the electron distribution is assumed to be isotropic. 
The entire description of the electron energy distribution is then given by three parameters:
the normalization, the peak Lorentz factor $\g^\prime_{pk}$, and the spectral curvature
parameter $b$.
Electrons with
$\gp \approx  \g^\prime_{pk}$ therefore contain the maximum nonthermal electron energy per 
unit logarithmic 
interval $\Delta(\log \gp)$ in the electron distribution.
Thus
$$\g^{\prime 2} N_e^\prime(\gp)=\gamma_{pk}^{\prime 2}N^\prime_e(\g^\prime_{pk})({\g^\prime\over\g^\prime_{pk}})^{-b\log({\g^\prime\over \gamma_{pk}^\prime})}$$
\begin{equation}
%\g^{\prime 2} N_e^\prime(\gp)=\gamma_{pk}^{\prime 2}N^\prime_e(\g^\prime_{pk})({\g^\prime\over\g^\prime_{pk}})^{-b\log({\g^\prime\over \gamma_{pk}^\prime})}
\equiv K y^{-b\log y} =  K y^{-\hat b\ln y} ,
\label{g2Ng1}
\end{equation}
where $y\equiv \gamma^\prime/\gamma^\prime_{pk}$, and $\hat b = b/\ln 10$. The electron distribution function is a smoothly curving function with continuously varying spectral index.
Normalizing to the total comoving nonthermal electron energy ${\cal E}^\prime_e$ through ${\cal E}^\prime_e = m_ec^2 \int_1^\infty d\gp\,\gp\,N_e^\prime (\gp )$ gives
${\cal E}^\prime_e \cong m_ec^2\gamma_{pk}^{\prime 2}\,N_e^\prime (\gamma_{pk}^\prime )I(b)$, so
\begin{equation}
K = {{\cal E}^\prime_e \over m_ec^2I(b)}\;,
\label{Kdef}
\end{equation}
 where $I(b) = \sqrt{\pi \ln 10/b}$ ($\cong 2.69/\sqrt{b}$). 
Note that the relationship is not exact, because we have replaced $1/\gamma_{pk}^\prime$ with 0 in the lower limit of the integral { (see Appendix for error estimate
and Section 6.3 for more detail on the log-parabola function and particle acceleration.)}.\footnote{This solution neglects a larger class of solutions that are not symmetric in $y\rightarrow 1/y$. 
This deficiency can be partly mitigated 
by using separate low-energy ($y\leq 1$) and high-energy ($y> 1$) width parameters. We do the symmetrical case here, reserving the 
mixed case for separate study.}

%The log-parabola function,  { originally introduced
%because it gave a good fit to the curving synchrotron SEDs of ``active" (blazar-like) radio sources
%\citep{1986ApJ...308...78L} and to the $\gamma$-ray spectra
%of Mrk 421 and Mrk 501 \citep{1999ApJ...511..149K}, has since been used to fit 
%blazar \citep{2002babs.conf...63G,2004A&A...413..489M} 
%and gamma-ray burst \citep{2010ApJ...714L.299M} spectra.}   It has also been recognized as 
%a convenient functional form for the electron energy distribution \citep{2006A&A...448..861M}.
%\citet{2011ApJ...739...66T} examine the properties of this function in detail, and 
%show how this form can arise in a second-order acceleration scenario.

%%%%%%%%%%%%%%%%%%%%%%%%%%%%%%%%%%%%%%%%%%%%%%%%%%%%%%%%%%%%%%%%%%%%%%
%%%%%%%%%%%%%%%%%%%%%%%%%%%%%%%%%%%%%%%%%%%%%%%%%%%%%%%%%%%%%%%%%%%%%%
\section{Equipartition Relations} \label{sec:equi}
%%%%%%%%%%%%%%%%%%%%%%%%%%%%%%%%%%%%%%%%%%%%%%%%%%%%%%%%%%%%%%%%%%%%%%
%%%%%%%%%%%%%%%%%%%%%%%%%%%%%%%%%%%%%%%%%%%%%%%%%%%%%%%%%%%%%%%%%%%%%%

The principle of equipartition is attractive by providing
a minimum power solution on the emissions from the jets of blazars that 
already demand extreme powers and large energies. In the black-hole 
jet system, the particle-field relations depend on uncertain 
jet-formation, particle acceleration and radiation 
mechanisms, for example, whether the jet power is provided by accretion
or black-hole rotation, and whether the jet plasma is particle or field 
dominated. 
%and furthermore gives the most economical solution. 
We avoid poorly understood microphysics by assuming
that a condition of equipartition holds between 
magnetic-field and nonthermal electron energy densities, 
as used in the analysis of a wide variety of 
astrophysical systems; see  \citet{pac70} for a
study of radio sources,
and \citet{bk05} for a more recent review.
Note that there is large uncertainty in the ``equipartition magnetic field"  if it 
is instead coupled to the  poorly known baryon content of the system.  
 In practical terms, the principle of equipartition
limits the analysis to a manageable scope 
by eliminating a parameter connecting
the magnetic-field and lepton energy densities.

One goal is to determine, by comparison with blazar SED data, whether the 
synchrotron and $\gamma$-ray spectra are reasonably well fit by
a log-parabola function for the electron energy distribution,
within the scenario that the blazar jet is described by a standard 
relativistic jet model with quasi-isotropic
electrons trapped in a randomly oriented magnetic field with 
mean strength $B^\prime$ in the fluid frame.  { The peak comoving electron 
Lorentz factor $\gamma_{pk}^\prime$ may represent the minimum Lorentz factor 
in a fast-cooling blast-wave scenario or the cooling Lorentz factor in a 
slow-cooling scenario \citep{1998ApJ...497L..17S},   or the value obtained by balancing 
acceleration and radiative cooling.} See also \citet{1996A&AS..120C.503G}.

\subsection{Parameters and derived quantities}

In the stationary (source) frame of the black-hole/jet system, we can list the important
observables that characterize the blazar SED:
\begin{enumerate}
  \item $\nu L_\nu^{pk,syn}$, the maximum value of the nonthermal synchrotron $\nu L_\nu$ spectrum. For a log-parabola synchrotron spectrum,
 the apparent isotropic bolometric synchrotron luminosity $L_{syn}=10^{48}L_{48}$ erg s$^{-1}$ is related to $\nu L_\nu^{pk,syn}$ by Equation (\ref{Lsyn});
  \item $\e_s= h\nu_s/m_ec^2 = 8.09\times 10^{-7}\nu_{14}$, the peak synchrotron frequency, that is, the frequency at which the $\nu L_\nu$ synchrotron spectrum reaches its maximum value, in units of $m_ec^2$, where $10^{14}\nu_{14}$ Hz is the peak synchrotron frequency in the source frame;
 % (thus the observed peak synchrotron frequency $\nu_s^{obs} = \nu_s/(1+z)$)
  \item $t_{var}= 10^4t_4$ s, the variability time scale, where $t_{var} = t_v^{obs}/(1+z)$, and $t_v^{obs}$ is the minimum 
  measured variability time scale, and we implicitly assume a roughly cospatial origin of the nonthermal radiation components (one-zone approximation);
{  \item $\nu L_\nu^{pk,C}$, the maximum value of the nonthermal Compton $\nu L_\nu$ spectrum. The Compton-dominance factor 
$A_{C} \cong  \nu L_\nu^{pk,{\rm C}}/\nu L_\nu^{pk,syn}\cong L_{\rm C}/L_{syn}$, gives the ratio of the $\gamma$-ray and synchrotron luminosities, assuming that the $\gamma$ rays are  from Compton scattering by the same jet electrons making the synchrotron emission; and
  \item $\e_c= h\nu_c/m_ec^2$, where the  peak Compton frequency $\nu_c$ is the frequency at which the $\nu L_\nu$ Compton spectrum reaches its maximum value. }
 % \item $A_{SSC} \equiv L_{SSC}/L_{s}$, the internal, or self-Compton-dominance factor, giving the ratio of the SSC luminosity to $L_s$. This ratio is more difficult to determine from the blazar SED, because the SSC component is concealed by the EC component, though it may be revealed at hard X-ray energies.
\end{enumerate}
A complete SED contains additional information in the detailed features of the spectrum.
{ In particular, the Compton component in blazars is generally a composite of the SSC 
and EC components, the latter of which depends on the external radiation environment. IR from 
thermal dust, the stellar content of the host galaxy, and 
hot accretion-disk radiation
may all contribute to the IR -- UV radiation spectrum, and must be considered when fitting data.}

Within the framework we set up, the derived synchrotron/SSC/EC model parameters are:
\begin{enumerate}
\item $B^\prime$, the comoving magnetic field, in G;
\item $\delta_{\rm D}$, the Doppler factor; 
\item $\gamma^\prime_{pk}$, the peak electron Lorentz factor;
\item $r_b^\prime = c\dD t_{var}$, the comoving radius of the emission region;
\item $u_i$, the external radiation field energy density; and
\item $\e_i=  h\nu_i/m_ec^2$, the dimensionless external radiation photon frequency,
\end{enumerate}
{ with $i$ indexing the various radiation fields in the blazar's environment.
In standard modeling approaches, $B^\prime$, $\delta_{\rm D}$ and $u_i$ are adjusted
to obtain a good fit, whereas in this approach, the equipartition assumption determines
$B^\prime$, $\delta_{\rm D}$, and $\gamma^\prime_{pk}$, assuming that the equipartition condition 
is valid, while  spectral fitting implies $u_i$. 
} 
%Note that parameter $k$ is formally included in the relation between $r_b^\prime$ and $t_v$.
%The term $k$ only appears in the combination $k t_v$, so that no additional parameter 
%is introduced if the value of $t_v$ is allowed to vary. 

The dynamical crossing time associated with 
the Schwarzschild radius of a $10^9 M_\odot$ black hole is $\approx 10^4$ s. 
Here we let $t_4 = 1$ in most of our calculations, noting that this parameter 
should be determined from    blazar variability  measurements at the time 
that the observing campaign took place.

The comoving magnetic field energy density $u^\prime_{B^\prime}= B^{\prime 2}/8\pi$.
In the standard blazar model, we can write the total comoving
energy density $u^\prime_{tot}$ of 
particles, field, and radiation in the 
outflowing plasma blob through the expression
\begin{equation}
u^\prime_{tot} = u^\prime_{B^\prime} + u^\prime_{par} +u_{ph}^\prime = u^\prime_{B^\prime} (1 +\zeta_{par} + \zeta_{ph}).
\label{u'tot}
\end{equation}
The subscripts refer to magnetic field ($B^\prime$), both large-scale ordered and disordered fields, including 
MHD waves; particles (par), divided into electronic (e) and nuclear (p/nuc) components (thus $\zeta_{par} = \zeta_e + \zeta_{p,nuc}$); 
and photons (ph). { We separately calculate the particle and field power, on the one hand, and the photon power on the other.}
The photon fields consist of the internal synchrotron (syn) and SSC fields, and the Compton-scattered $\gamma$ rays. 

%{ The equipartition relations derived below depend on the internal synchrotron energy 
%density, $u^\prime_{syn}$, which can be determined from the SSC luminosity $L_{SSC}$,
%or the self-Compton dominance factor $A_{SSC} = L_{SSC}/L_{syn} \approx u^\prime_{syn}/u^\prime_{B^\prime}
%\equiv \zeta_s$ (the inequality is exact in the Thomson regime). Likewise
%for the external Compton parameters, $A_{EC} = L_{EC}/L_{syn} = u^\prime_*/u^\prime_{B^\prime}\approx \zeta_{*}$.
%Fitting to the hard X-ray and $\gamma$-ray spectrum yields $\zeta_s$ and $\zeta_{*}$.
% }
 
The simplest equipartition relation is $u^\prime_e = \zeta_e u^\prime_{B^\prime}, \zeta_e \cong 1$, which also minimizes jet power.\footnote{ 
The particle energy density is 4/3rd the magnetic-field energy density at minimum jet power; 
see \citep[see, e.g.,][]{2001MNRAS.327..739G,2008ApJ...686..181F}. For simplicity, we let 
$\zeta_e = 1$ and $\zeta_{p,nuc} = 1$ in the calculations. 
%This does not affect power calculations, which are dominated by photon power.
}
We also let $\zeta_{p,nuc} = 1$ in the following calculations, representing a jet with low baryon
content. The baryonic content affects only the total power and not spectral properties; see Section 6.4.
%return later to the impact of letting $\zeta_{p,nuc} \sim 10$ -- 100.)   

%For the SSC component in HSP BL Lac objects, a modest range of $A_{SSC}$ is implied by the multiwavelength SEDs, though 
%Klein-Nishina effects also play a role. Observations of the SEDs of blazars show a wide range of Compton dominance
%parameters, which probably relates to the energy densities of the 
%external radiation field that can vary widely, depending on the environment and
%location of the emission region. 

%%%%%%%%%%%%%%%%%%%%%%%%%%%%%%%%%%%%%%%%%%%%%%%%%%%%%%%%%%%%%%%%%%%%%%
%%%%%%%%%%%%%%%%%%%%%%%%%%%%%%%%%%%%%%%%%%%%%%%%%%%%%%%%%%%%%%%%%%%%%%
\subsection{Constraints and equipartition relations} \label{sec:constraints}
%%%%%%%%%%%%%%%%%%%%%%%%%%%%%%%%%%%%%%%%%%%%%%%%%%%%%%%%%%%%%%%%%%%%%%
%%%%%%%%%%%%%%%%%%%%%%%%%%%%%%%%%%%%%%%%%%%%%%%%%%%%%%%%%%%%%%%%%%%%%%

One advantage of using the log-parabola function, besides having only 
three parameters to characterize the nonthermal electron distribution, is
that in the limit $b\rightarrow \infty$, the electrons become  
monoenergetic. The solution, along with 
corrections
 related to the width of the log-parabola electron distribution, is
derived in the Appendix. The $b\rightarrow \infty$ limit
is obtained by letting $f_1\rightarrow 1$ and $ f_2 \rightarrow 1$ in the equations below; see Table \ref{table0}).
The analytic results provide parameters as input for the numerical model.

\begin{table}[t]
\begin{center}
\caption{Log-Parabola and Geometry Correction Factors}
\begin{tabular}{cl}
\hline
% &   &   & n  &  & $\nu$ &    & $\nu_\beta$ &  \\
$I(b)$ & $\sqrt{\pi\ln 10/b}$  \\ 
$f_0$ & $1/3$  \\ 
$f_1$ & $10^{-1/4b}$  \\ 
$f_2$ & $10^{1/b}$  \\ 
$f_3$ & ${f_1\over 2I(b)}= {10^{-1/4b}\over 2 \sqrt{\pi\ln 10/b}}$ \\ 
\hline\end{tabular}
\label{table0}
\end{center}
\vskip0.2in
\end{table}

A first constraint arises from simple kinematics. In a blob geometry,
the comoving synchrotron energy density $u^\prime_{syn} = L_{syn}/ 4\pi r_b^{\prime 2} c \delta_{\rm D}^4f_0$,
where $r_b^\prime \cong c\dD t_{var}$, so 
\begin{equation}
%L^{kin}_s =  {4\pi r_b^{\prime 2} c \delta_{\rm D}^4 u^\prime_s  }\;.
L_{syn} =  4\pi t_{var}^2 c^3 \delta_{\rm D}^6 u^\prime_{syn}  f_0\;,
\label{uprimes2}
\end{equation}
where $f_0$ is a factor of order unity,\footnote{For a spherical blob geometry,
 $f_0 \cong 1/3$. For a spherical shell geometry, $f_0\cong 1$  and $\delta_{\rm D} \rightarrow \Gamma$.} and
$u^\prime_{syn} \equiv \zeta_s u^\prime_{B^\prime}$ relates the internal synchrotron
 and magnetic fields.
%The apparent bolometric synchrotron luminosity $L_s$ is an observable, 
%provided the redshift is known.
A second constraint
%, which is exact in the limit $b\rightarrow \infty$, 
is  the synchrotron luminosity constraint 
\begin{equation}
L_{syn} = {4\over 3}\,c\sigma_{\rm T} \,u_{B^\prime}^\prime N_{e0} \gamma_{pk}^{\prime 2}\delta_{\rm D}^4\;,
\label{Ls}
\end{equation}
which takes this form even with log-parabola width corrections, as shown in the Appendix, Equation (\ref{Lsyncorr}).
When the equipartition relation 
\begin{equation}
u_e^\prime = {{\cal E}_e^\prime\over V_b^\prime} = {m_ec^2 N_{e0} \gamma_{pk}^\prime\over V_b^\prime}f_1  = \zeta_eu_{B^\prime}^\prime\;,
\label{ueprime}
\end{equation}
with $V_b^\prime = 4\pi r_b^{\prime 3}/3$  and $ f_1 \equiv  10^{-1/4b}$ (see Appendix, Equation (\ref{Ee0})), is substituted into the synchrotron constraint, Equation (\ref{Ls}), 
and the result is equated with $L_{syn}$ given by Equation (\ref{uprimes2}), we find
\begin{equation}
\zeta_s = {4\over 9}\sigma_{\rm T}u_{B^\prime}^\prime{r_b^\prime\over m_ec^2 f_0 f_1}\zeta_e \gamma_{pk}^\prime\;.
\label{uprimes}
\end{equation}

%In the $b \rightarrow \infty$ asymptote, 
The peak synchrotron frequency can be expressed in terms
of the critical magnetic field $B_{cr} = m_e^2c^3/ e\hbar$ as
\begin{equation}
\e_s = f_2\epsilon_{pk} = f_2 \delta_{\rm D}\epsilon^\prime_{pk} = f_2\delta_{\rm D}{3\over 2}\,{B^\prime \over B_{cr}} \gamma_{pk}^{\prime 2}\;,
\label{es}
\end{equation}
{where $f_2 \equiv 10^{1/b}$}, from which, with Equation (\ref{uprimes}), we derive
\begin{equation}
\delta_{\rm D}B^{\prime 3} = {3f_2\over 2B_{cr}\e_s}\,\left({18\pi m_e c f_0f_1\zeta_s\over \sigma_{\rm T}t_{var}\zeta_e}\right)^2\;.
\label{delta}
\end{equation}
Replacing this result in eq.\ (\ref{uprimes2}) gives
$$\delta_{\rm D} = \left( {2 L_{syn}^3\over 3^{10} \pi^4 f_0^7 \zeta_s^7 c^{13}}\right)^{1/16}\left( {\sigma_{\rm T} \zeta_e\over f_1 m_e}\right)^{1/4}
\left({B_{cr}\epsilon_s\over f_2t_{var}}\right)^{1/8}$$
\begin{equation}
\cong 
17.5 L_{48}^{3/16}{\zeta_e^{1/4} \over \zeta_s^{7/16}}{\nu_{14}^{1/8}\over t_4^{1/8}}\;{1\over f_0^{7/16} f_1^{1/4} f_2^{1/8}}\;.
\label{deltaD}
\end{equation}
%Note that  $\nu L_\nu^{pk,syn}$ is related to $L_{syn}$ through $L_{syn} = 2\sqrt{\pi \ln 10/b}\; \nu L_\nu^{pk,syn}$ (Equation (\ref{}).
Inverting Equation (\ref{es}) to get an expression for $B^\prime$ as a function of $\delta_{\rm D}$, and using Equation (\ref{deltaD}) for $\delta_{\rm D}$, 
gives  
\begin{equation}
B^{\prime}({\rm G}) \cong {5.0 \,\zeta_s^{13/16}\over L_{48}^{1/16}t_4^{5/8} \nu_{14}^{3/8}\zeta_e^{3/4}}\;f_0^{13/16} f_1^{3/4} f_2^{3/8}\;.
\label{BprimeG}
\end{equation}
 Using eqs.\ (\ref{es}), (\ref{deltaD}), and (\ref{BprimeG}), we have 
\begin{equation}
\gamma_{pk}^\prime \cong 523\;{\nu_{14}^{5/8}\zeta_e^{1/4}\over L_{48}^{1/16} \zeta_s^{3/16} }\,t_4^{3/8}\;{1\over f_0^{3/16} f_1^{1/4} f_2^{5/8}}\;.
\label{gammapprime}
\end{equation}

\subsection{EC component}

In the Thomson regime, the peak frequency of the external Compton-scattered component is
%$$\e_{\rm T}^{EC} \cong $$
\begin{equation}
%\e_{\rm T}^{EC} \cong f_2\,{4\over 3}\,(\delta_{\rm D}\gamma_{pk}^\prime)^2 \,\e_*\cong
\e_{\rm T}^{EC} \cong  f_2\,{4\over 3}\,(\delta_{\rm D}\gamma_{pk}^\prime)^2 \,\e_* \cong 
2\times 10^3 \,{\epsilon_*\over 2\times 10^{-5}}{\zeta_e\over \zeta_s^{5/4}}\,{\nu_{14}^{3/2} t_4^{1/2} L_{48}^{1/4}\over f_0^{5/4} f_1 f_2^{1/2}}\;,
%{f_2\zeta_e\over \zeta_s^{5/4}}\,\nu_{14}^{3/2}L_{48}^{1/4}\sqrt{t_4}\,\e_{\rm Ly\alpha}\;.
\label{eCEC}
\end{equation}
{ normalizing to the Ly $\alpha$ energy $\epsilon_* \cong 2\times 10^{-5}$ for the mean photon energy in the external radiation field.}
The onset of the Klein-Nishina correction for external Compton scattering takes place when $4\gamma_{pk}^\prime \e^\prime\gtrsim 1$. Taking $\e^\prime \cong \delta_{\rm D}\e_*$, the photon energy in the stationary frame where KN effects become important is at $\e_{\rm C,KN} = 1/12\e_*$. For Ly$\alpha$ target photons,
$\epsilon_{Ly\alpha } = 1$,
% photons with $\e_* = 2\times 10^{-5}$ $(\epsilon_{{\rm Ly}\alpha}= 1)$, 
$E_{\rm C,KN}\cong 2.2$ GeV, and this could be the reason for the  spectral cutoff in FSRQs and low-synchrotron peaked (LSP; $\nu_s^{obs} < 10^{14}$ Hz) blazars \citep{2010ApJ...721.1383A,cer13}. 
{ The strongest dependence of the peak frequency of the external Compton component is $\e_{\rm T}^{EC}\propto \nu_{14}^{3/2}$, so that most
FSRQs with $\nu_{14}\cong 0.1$ have $\e_{\rm T}^{EC} < \e_{\rm C,KN}$, and the combined effects of the softening electron spectrum and
Klein-Nishina cross section make an apparent cutoff at $\approx \e_{\rm C,KN}$.}

\subsection{SSC component}

The peak $\gamma$-ray energy in the Thomson approximation of the SSC spectrum is approximately given by 
$$\e_{\rm T}^{SSC} \cong f_2\,{4\over 3} \dD \gamma_{pk}^{\prime 2} \e_s^\prime \cong f_2\,{4\over 3} \gamma_{pk}^{\prime 2} \e_s\;$$
\begin{equation}
\cong {9.4\times 10^3 \over f_0^{3/8} f_1^{1/2} f_2^{1/4}}\,{\zeta_e^{1/2}\over \zeta_s^{3/8}}\,{t_4^{3/4}\,\nu_{16}^{9/4}\over 
L_{48}^{1/8}}\;,
\label{eC}
\end{equation}
defining $\nu_{16} = \nu_s/10^{16}$ Hz. Compared with the Thomson approximation, 
Klein-Nishina effects in SSC scattering are important for SSC $\gamma$ rays with 
energies $\epsilon^{SSC}_{\rm KN}\cong (4/3) f_2 \gamma^{\prime 2}_{pk}\epsilon_s\cong 
 f_2 \gamma_{pk}^{\prime } \delta_{\rm D}/3$. Thus
%$$\epsilon_{\rm KN} \cong 9400\;{\nu_{16}^{9/4} t_4^{3/4}\over L_{48}^{1/8}}\,{ \zeta_e \over f_0^{3/8} f_1^{1/2} f_2^{1/4} \zeta_s^{3/8}}$$
\begin{equation}
%\epsilon_{\rm KN} \cong 9400\;{\nu_{16}^{9/4} t_4^{3/4}\over L_{48}^{1/8}}\,{f_2 \zeta_e \over f_0^{3/8} f_1^{1/2} f_2^{5/4} \zeta_s^{3/8}}
\epsilon_{\rm KN}\cong 3050\;{\nu_{14}^{3/4} t_4^{1/4} L_{48}^{1/8}}\,{f_2^{1/4} \zeta_e^{1/2} \over f_0^{5/8} f_1^{1/2}  \zeta_s^{5/8}}\;.
\label{epsilonKN}
\end{equation}
For a given parameter set,  KN effects are important 
when $\e_s\gtrsim \e_{s,{\rm KN}} = \dD/4\gamma_{pk}^\prime$, that is,  
for synchrotron photons with energies greater  than $\e_{s,{\rm KN}}$, where 
\begin{equation}
\e_{s,{\rm KN}} \cong 8.4\times 10^{-3}\;\sqrt{ {\sqrt{f_2L_{48}/f_0\zeta_s}\over f_1 \nu_{14}t_4 }
}\;,
\label{esKN}
\end{equation}
or
\begin{equation}
\nu_{14,{\rm KN}} \cong {470\over (f_1 t_4)^{1/3} }\;\left({f_2L_{48}\over f_0 \zeta_s}\right)^{1/6}\;.
\label{esKN}
\end{equation}
Thus, KN effects are most important in high-synchrotron peaked (HSP; $\nu_s^{obs} > 10^{15}$ Hz) BL Lac objects.

\subsection{Synchrotron self-absorption}

We use the $\delta$-function approximation to calculate synchrotron self-absorption (SSA). 
The SSA coefficient for a photon of frequency $m_ec^2\ep/h$ in the fluid frame is given in this approximation by
\begin{equation}
\kappa_{\ep} = {-\pi \over 36}\,{\lambda_{\rm C} r_e\over \ep}\,\left[\gamma^\prime_s\,{\partial \over \partial \gamma^\prime_s} \left( {n_e^\prime(\gamma^\prime_s )\over \gamma_s^{\prime 2}}\right)\right]\;,\;{\gamma^\prime_s = \sqrt{\ep/2\varepsilon_{B^\prime} } }\;,
\label{kappaepsilon}
\end{equation}
\citep[][eq.\ (7.144), with corrections]{dm09}.
Here $\lambda_{\rm C} = h/m_e c$ is the Compton wavelength, $r_e = e^2/m_ec^2$ is the classical electron radius, and 
$\varepsilon_{B^\prime} = B^\prime/B_{cr}$.
Taking $n_e^\prime(\gamma^\prime ) = N^\prime_e(\gp )/V_b^\prime$, 
%with $V_b^\prime = 4\pi r_b^{\prime 3}/3$, 
and substituting
Equation (\ref{g2Ng1}) for $N_e^\prime(\gp )$ gives
\begin{equation}
\kappa_{\ep} = {\pi \over 18}\,{{\cal E}_e^\prime \over m_ec^2 I(b) V_b^\prime \gamma_{pk}^{\prime 4}}\, 
{\lambda_{\rm C} r_e\over \ep} (2+ b\log \hat y )\,\hat y^{-(4+b\log \hat y)}\;,
\label{kappaepsilon1}
\end{equation}
where $\hat y \equiv \sqrt{\ep/2\varepsilon_{B^\prime}}/\gamma_{pk}^\prime$.

The SSA opacity $\tau_{\ep} = 2\kappa_{\ep} r_b^\prime$. The unabsorbed spectrum is multiplied
by the factor $u(\tau ) = 1/2 + \exp(-\tau )/\tau - [1-\exp(-\tau )]/\tau^2 $, with $\tau =\tau_{\ep}$, 
to give the absorbed  spectrum \citep{1979A&A....76..306G}.

\subsection{Jet power}

We calculate absolute jet power $L_{jet}$ for a two-sided jet using the relation
\begin{equation}
L_{jet} = L_{B,par}+L_{ph} = 2\pi r_b^{\prime 2} \beta \Gamma^2  c (u_{B^\prime}^\prime + u_{par}^\prime)+L_{ph}
\label{Ljet}
\end{equation}
 \citep[][compare  Equation (\ref{u'tot})]{1993MNRAS.264..228C,2008MNRAS.385..283C}.
The absolute photon power $L_{ph} =L_{abs,s} + L_{abs,{\rm C}}$ 
is separately evaluated from the magnetic-field and particle power, $L_{B,par}$,
because the photon distributions in the comoving frame can have peculiar 
anisotropies associated with the
different angular distributions of internal and external Compton scattering
\citep{1995ApJ...446L..63D}. The apparent $L_{syn}$ and absolute $L_{abs,s}$ synchrotron 
and SSC fluxes are related by the expression
\begin{equation}
L_{abs,syn} = 8\Gamma^2 L_{syn}/3\delta_{\rm D}^4\;.
\label{Ljetsyn}
\end{equation}
For external Compton (C) emission, the relationship is 
\begin{equation}L_{abs,{\rm C}} = 32\Gamma^4 L_{{\rm C}}/5\delta_{\rm D}^6\;
\label{LjetEC}
\end{equation}
\citep[see Appendix A in][]{2012ApJ...755..147D}.
%and 
%\begin{equation}L_{ph} = L_{abs,s} + L_{abs,{\rm C}} \;.
%\label{Lph}
%\end{equation}

These very different behaviors are important for power requirements in jets.
The kinetic power, $L_{B^\prime,par} = (1+\zeta_{e}+\zeta_{p,nuc})\Gamma u_{B}^\prime V_b^\prime/t_{var}$, 
accounts for the magnetic-field and particle energy in the jet. In our model calculations,
the magnetic-field and particle power is dominated by the photon power $L_{ph}$, so that baryon-loading does 
not have a big impact on jet power requirements until $\zeta_{p,nuc}\gg 1$. In actual fitting
to 3C 279, we find that $L_{B,par} \approx L_{ph}$.

%%%%%%%%%%%%%%%%%%%%%%%%%%%%%%%%%%%%%%%%%%%%%%%%%%%%%%%%%%%%%%%%%%%%%%
%%%%%%%%%%%%%%%%%%%%%%%%%%%%%%%%%%%%%%%%%%%%%%%%%%%%%%%%%%%%%%%%%%%%%%
\section{Numerical Results} \label{sec:numerical}
%%%%%%%%%%%%%%%%%%%%%%%%%%%%%%%%%%%%%%%%%%%%%%%%%%%%%%%%%%%%%%%%%%%%%%
%%%%%%%%%%%%%%%%%%%%%%%%%%%%%%%%%%%%%%%%%%%%%%%%%%%%%%%%%%%%%%%%%%%%%%

Using the derived values of $\dD$, $B^\prime$, $\g^\prime_{pk}$, and $r_b^\prime$, 
along with the log-parabola description of the electron spectrum,
Equation (\ref{g2Ng}), we numerically calculate 
$\nu L_\nu$ SEDs. The method for calculating the SEDs are 
presented in \citet{2008ApJ...686..181F} and \cite{2009ApJ...692...32D}.
Assuming that the external radiation fields are locally isotropic,  
the { assumed isotropic  distribution of electrons in the jet frame is transformed
to the source frame through the expression $N(\gamma,\Omega) = \delta_{\rm D}^3 N^\prime(\gamma^\prime)/4\pi$, so 
all angular dependence of the particle distribution in the 
stationary black-hole frame from the small spherical jet blob is in the Doppler factor \citep{2001ApJ...561..111G}.
The spectrum of external Compton $\gamma$ radiation is calculated 
assuming the head-on approximation for the Compton cross section, which applies
to all but $\sim 1/\gamma^2$ of the photons for scattering by electrons
with Lorentz factor $\gamma$. Furthermore, 
the scattered photon is assumed to travel in the direction of the 
incident electron, which holds when $\epsilon \ll \gamma$.  
For the head-on approximation  of the Compton cross section in the general case of anisotropic 
and isotropic external radiation fields, see eqs.\ (6.97) and eq.\ (6.113) in \citet{dm09}, respectively. The 
isotropic case, assumed here, recovers the \citet{1968PhRv..167.1159J} result.}
%This approach is valid for relativistic electrons ($\gamma\gg 1$) scattering isotropic lower-energy photons ($\epsilon\ll\gamma$), 
%with Klein-Nishina effects fully taken into  account. 

In the model calculations shown
in this section, we calculate SEDs for external isotropic photon sources with $\epsilon_* = 2\times 10^{-5}$
characterizing the BLR ($10.2$ eV), and $\epsilon_* = 5.4\times 10^{-7}$ ($\cong 0.3$ eV) describing radiation from a hot dust torus
at 1200 K temperature. The 
former value is motivated by the strong Ly$\alpha$ line in blazars, specifically, 3C 454.3 \citep{2011MNRAS.410..368B}.
The latter value is motivated by observations of 4C +21.35 by \citet{2011ApJ...732..116M} 
who, using 5 -- 35$\mu$ Spitzer, Sloan Digital Sky Survey, Two
Micron All Sky Survey and Swift UVOT data, decompose the spectrum of 4C +21.35 
into a nonthermal power law and a two-temperature dust model. The warm dust component
has effective temperature $T\sim 660$ K and radiates $\approx 10^{45}$ erg s$^{-1}$, 
while the hot dust component has
$T\approx 1200$ K and with luminosity  $7.9\pm 0.2 \times 10^{45}$ erg s$^{-1}$.
Calculations of AGN emissions reprocessed by dust clouds \citep{2008ApJ...685..147N,2008ApJ...685..160N}
show that the peak power of the reprocessed emission occurs at $\approx 10\mu$ (the $10\mu$ silicate feature), 
with broad wings from $\approx 3\mu$ to $\sim 30\mu$. 
This peak frequency corresponds to $\epsilon_* \approx 2\times 10^{-7}$, or an effective blackbody temperature of $\approx 500$ K, 
characteristic of warm dust.

%Relevant for characterizing the IR environment of blazars are recent observations 
%of the unusual FSRQ 4C +21.35 by 
%\citet{2011ApJ...732..116M}. 
%We examine hot dust fits for 3C 279 below, but consider warm dust ($\epsilon_* = 2\times 10^{-7}$)
%for an initial parameter study.

When fitting actual blazar spectra, both IR dust and BLR photon sources contribute and have 
to be appropriately adjusted to give a good fit. Detailed fitting requires a 
range of atomic lines and multiple thermal components \citep{cer13}, but this two-line simplification
illustrates the technique.

\begin{table}[t]
\begin{center}
\caption{Blazar Model input$^a$}
\begin{tabular}{ccllllll}
\hline
% &   &   & n  &  & $\nu$ &    & $\nu_\beta$ &  \\
$ {\rm Case}$ & $\nu L_\nu^{pk,syn}$ & $\nu_{14}$ & $t_4$ &  $b$ & $\zeta_s$ &  $\zeta_*$$^b$ & $N_\Gamma$ \\ 
$$ & $(10^{48} erg/s)$ & $ $   &   &   &   &    &   \\ 
\hline
\hline
 &  &  &   &   &   &  &   \\ 
1a & 0.1 & 0.1 & 1 & 1 & 1 & 10 & 1\\
% &  &  &  &  & &   & &   \\ 
1b &  &    &  &  &  &  & 0\\
% &  &  &   &  &  &   & &  \\ 
%1c &    & &  &  &  & & \\
% &  &  &   &  & &   & &  \\ 
%1d &   &    &  &  &  &  & 1\\
 &  &  &     & &   & &  \\ 
2a & 0.01 &  0.1 & 1 & 1 & 1 &  & 1\\
% &  &  &  &  & &   & &   \\ 
2b &  &  &   &  &  &  & 0\\
% &  &  &   &  &  &   & &  \\ 
2c &   &  &  & 0.5 &  & & 1\\
% &  &  &   &  & &   & &  \\ 
2d &   &    &  & 2 &  &  & \\
% &  &  &   &  & &   & &  \\ 
2e &   &    & 0.1 & 1 &  &  & \\
% &  &  &     & &   & &  \\ 
2f &   &    & 10 &  &  &  & \\
 &  &  &     & &   & &  \\ 
3a & 0.01  & 0.1 & 1 & &  &   & \\
%3a2 &   &  &  &  &  &  & \\
3b &    &1 &  &  &  & & \\
%3b2 &     &  &  &  &  &  & \\
3c &     &10 &  &  &  & & \\
%3c2 &     &  &  &  &  &  & \\
 &  &    &  & &   & &  \\ 
4a$^c$ & 0.001  & $1$ &  &  &  & ~-- & \\
4b &  &   $10^1$ &  &  &  &   & \\
4c &   &    $10^2$ &  &  &    \\
4d &   &   $10^3$ &  &  &  & & \\
4e &   &   $10^4$ &  &  &  &  & \\
\\
\hline\end{tabular}
\label{table1}
\end{center}
$^a$ Soft photon energies are $\epsilon = 2\times 10^{-5}$, corresponding to Ly$\alpha$ radiation 
in the BLR, and $5.4\times 10^{-7}$, corresponding to 1200 K dust, with $\zeta_e  =\zeta_{p,nuc} =  1$. Blank entries in Tables are equal to first filled vertical entry above. { For Cases 4a-e, $\zeta_* \rightarrow 0$}.\\
$^b$ $\zeta_* = 10$ for each of the BLR and IR dust radiation fields\\
$^c$ Equipartition synchrotron/SSC model without external radiation fields
\vskip0.2in
\end{table}

\begin{table}[t]
\begin{center}
\caption{Blazar Model Output}
\begin{tabular}{cccccccc}
\hline
% &   &   & n  &  & $\nu$ &    & $\nu_\beta$ &  \\
${\rm Case}$ & $\Gamma$ & $\dD$ & $B^\prime$ & $\gamma^\prime_{pk}$ &  $u_{0}$   &   $r^\prime_b$ &  
$\log[L_{B,par}$ \\ 
 &   &   &  (G) &  & (${\rm erg}$ &  ($10^{15}$  & $(L_{ph})$ \\
 &   &   &   &  & ${\rm cm}^{-3}$)  &  cm)  & (erg s$^{-1}$)]$^a$ \\
\hline
\hline
 &  &  &   &   &  &   & \\ 

1a & 16.4 & 16.4 & 7.8 & 43.4 & 0.067    &4.9 & $45.9(47.2)$  \\
1b & 8.2 &  &  &  & 0.27 & &  $45.3(46.6)$  \\
%1c &  &  &  &  & & &  46.0(46.8)  \\
%1d & 17.8 &  &  &  & 0.0431 &  & $46.6(47.4)$  \\
 &  &  &   &   & & &   \\ 
2a & 10.6 & 10.6 & 9.0 & 50.0 & 0.21 &  3.2 & $45.3(46.6)$  \\
2b & 5.31 &  &  &  & 0.85 & &  $44.7(46.0)$ \\
2c & 9.8 & 9.8 & 13.5 & 44.9 & 0.57&   2.9 & $45.5(46.8)$\\
2d & 10.7 & 10.7 & 7.4 & 97.8 & 0.14 &  3.2 & $45.2(46.4)$  \\
2e & 14.2 & 14.2 & 38 & 21 &2.1 & 0.42 &  $45.1(46.3)$  \\
2f & 8.0 & 8.0 & 2.1 & 119 & 0.021 & 24 &  $45.6(46.8)$  \\
\\
3a & 10.6 & 10.6 & 9.00  & 50.0 & 0.21  & 3.2 & $45.3(46.6)$  \\
3b & 14.2 & 14.2 & 3.80 & 211 & 0.021  & 4.2 & $45.1(46.2)$  \\
3c & 18.9 & 18.9 & 1.60 & 890 & 0.0021  & 5.7 & $44.8(45.8)$  \\
\\
4a & 9.2 & 9.2 & 4.37 & 244 & -- & 2.8 & $44.4(44.4)$  \\
4b & 12.3 & 12.3 & 1.84 & 1030 &   &  3.7 & $44.2(44.2)$  \\
4c & 16.4 & 16.4 & 0.78 & 4340 &  &  4.9 & $43.9(43.9)$  \\
4d & 21.8 & 21.8 & 0.33 & 1.83e4 &   &  6.5 & $43.7(43.7)$  \\
4e & 29.1 & 29.1 & 0.14  & 7.71e5 &   &  8.7 & $43.4(43.2)$  \\
\\
\hline\end{tabular}
\label{table2}
\end{center}
$^a$$L_{B,par}$ is two-sided absolute magnetic-field and particle power, 
and $L_{ph}$ is photon power; see Equation (\ref{Ljet})
\vskip0.2in
\end{table}

\begin{figure}[t]
% \vspace*{-2.0 cm}
\begin{center}
 \includegraphics[width=3.5in]{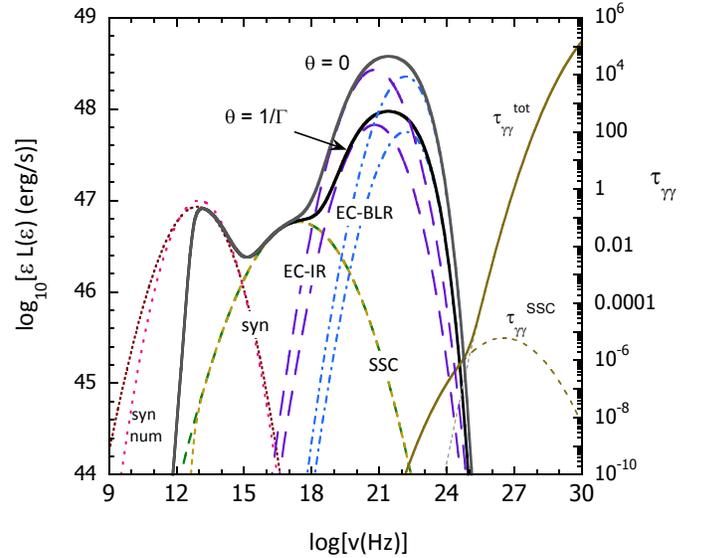} 
\vspace*{+0.2 cm}
 \caption{
Model blazar SEDs from radio to $\gamma$ rays implied by equipartition conditions,
using synchrotron luminosities and peak frequencies characteristic of a powerful FSRQ. 
In this calculation, $\zeta_* = 10$ for each of the BLR and IR components,
and $\zeta_s = 1$. 
The $\nu L_\nu$ peak synchrotron luminosity $\nu L_\nu^{pk,syn} = 10^{47}$ erg s$^{-1}$, 
and the bolometric synchrotron luminosity is a factor $2\sqrt{\pi\ln 10} \cong 5.38$ larger. 
Jets are viewed off axis, $\theta \cong 1/\Gamma$, and on-axis, $\theta = 0$, 
as labeled. The BLR target photons are approximated as  Ly$\alpha$ ($\e_* = 2\times 10^{-5}$),
 and the IR radiation as monochromatic photons with energy
 ($\e_* = 5.38\times 10^{-7}$). Separate synchrotron, SSC, and 
EC components are shown.  Analytic expressions for the synchrotron SED (``syn num") 
and $\gamma\gamma$ opacity through the synchrotron and SSC fields ($\tau_{\gamma\gamma}$) 
are also plotted.}
\label{fig1}
\end{center}
\end{figure}

Figure \ref{fig1} shows model SEDs for values characteristic of 
a powerful FSRQ blazar with a bolometric synchrotron luminosity $L_s \cong 5\times 10^{47}$ erg s$^{-1}$
and peak photon frequency $\nu_s = 10^{13}$ Hz. 
Input values for this and other figures in this section are given in Table \ref{table1}, and 
derived values are listed in Table \ref{table2}.  
In the calculations, we use the expression 
\beq \Gamma = {1\over 2}\,{(1 +N_\Gamma^2)\delta_{\rm D} }\;
\label{GammaDelta}
\eeq
to relate $\Gamma$ and $\delta_{\rm D}$, where $N_\Gamma \equiv \Gamma\theta$, which is
valid in the limit $\Gamma \gg 1$, $\theta\ll 1$. We model off-axis blazars with 
 $N_\Gamma = 1$, and on-axis blazars with $N_\Gamma = 0$.  The BLR is approximated by 
an external monochromatic radiation field with line energy $\epsilon_* = 10.2/511000 = 2\times 10^{-5}$,
 and likewise for the  IR torus with  mean photon energies characteristic
of 440~K and 1200~K blackbodies.

The different beaming factors of the synchrotron and SSC components on 
the one hand and the EC components on the other are seen when the 
$\nu L_\nu$ $\gamma$-ray SEDs for $\theta = 1/\Gamma $, $\delta_{\rm D} = \Gamma $ is compared with 
the SED for $\theta=0, \dD = 2 \Gamma$ \citep{1995ApJ...446L..63D,2001ApJ...561..111G}. 
In Figure \ref{fig1}, $\Gamma \approx 16$ for $\theta = 1/\Gamma $ and $\Gamma \approx 8$ for
$\theta = 0$, with  photon energy densities $u_0\approx 0.067$ erg cm$^{-3}$ and $u_0 = 0.27$
erg cm$^{-3}$  for the off- and on-axis cases, respectively.
A larger external energy density is required for the on-axis jet model when all other parameters are
the same, because the implied $\Gamma$ factor is smaller. The magnetic field $B^\prime \cong 8$ G
and $\gamma^\prime_{pk}\cong 43$.

For the equipartition solution, a break is formed at $\approx 400$ MeV -- GeV energies when jet electrons scatter
Ly$\alpha$/BLR photons \citep{cer13}. If IR photons alone are scattered, the $\nu L_\nu$ spectrum peaks near 10 MeV and 
the break is at higher energies, but below 1 GeV.
If the electrons in the flaring jet are within the BLR, they scatter both IR and BLR photons, whereas when 
the radiating jet is found far outside the BLR, scattering of IR photons dominates. So an FSRQ with a GeV break would also be accompanied by a $\gamma$-ray external Compton dust feature. In contrast, flares occurring far outside the BLR  would make a blazar SED peaked at MeV energies, as seen in some blazars
with low Compton peak frequencies, e.g., 
CTA 102 and PKS 0528+134
%, from analysis of Compton Observatory data 
\citep{1995ApJ...451..575M}.

The value $\zeta_s = 1$ in Figure \ref{fig1}, and the bolometric SSC luminosity is equal to the synchrotron luminosity $L_{syn}$ 
considering that the SSC SED has a larger width (or smaller effective value of $b$) than the synchrotron SED's width. Also shown here is 
{  a $\delta$-function expression for the synchrotron luminosity spectrum $\epsilon L_{syn}(\epsilon) 
=\upsilon x^{1-\hat b \ln x}$, derived in the Appendix, Equation (\ref{eLsyne}).
%using the $\delta$-function synchrotron relation  $\epsilon_{syn} = (3/2) \delta_{\rm D} (B^\prime/B_{cr})\gamma_{pk}^{\prime 2}$,
Here $x = \sqrt{\epsilon/\epsilon_{s}}$
and $\upsilon = c\sigma_{\rm T} B^{\prime 2}\delta_{\rm D}^4 \gamma_{pk}^\prime K/12\pi$; see Eq.\ (\ref{Kdef}).
The actual peak photon energy $\epsilon_s$ of the luminosity spectrum
is simply $\epsilon_s = \e_{pk}10^{1/b}$. 
%Thus for $b = 1$, $\hat\epsilon$ is an order of magnitude
%higher than the value of $\hat\epsilon$ implied by $b\rightarrow \infty$, as confirmed by the calculations. 
%Moreover, $\hat \epsilon L_{syn}(\hat\epsilon)
%= 10^{1/4b}\kappa $, and the ratio of bolometric to peak synchrotron luminosities is
% $\int_0^\infty d\epsilon L_{syn}(\epsilon)/ \hat \epsilon L_{syn}(\hat\epsilon)
%= 2\sqrt{\pi \ln 10/b}$. 
}

The SSC spectrum, using either the optically thin or the self-absorbed target synchrotron photon fields,
is shown for comparison in Fig. {\ref{fig1}. The effects of self-absorption on the SSC specrum are negligible.
%above the SSA frequency, which is  $\gtrsim 10^{12}$ Hz for this model. 
%For the same reason, the peak frequency 
%of the Compton-scattered dust component gets shifted to higher energies than expected from simple Thomson estimates.
%When $\theta \approx 1/\Gamma$, the Thomson-scattered $\gamma$-ray luminosity $L_\gamma$ of the dust component 
% is about $\zeta_* (= 10)\times$ larger than $L_s$, but is even larger when viewed down the jet axis 
%due to the different beaming factors. 
Because of the stronger Klein-Nishina effects for the scattered BLR photons, 
 the Compton-scattered Ly$\alpha$ component has a smaller apparent 
luminosity than that of the dust component. 
 The absolute jet luminosities for these cases almost reach the Eddington luminosity for a $10^{9}M_\odot$ black hole when $\theta= 1/\Gamma$, but are significantly less 
if viewed on-axis.
The internal $\gamma\gamma$ opacity of a photon through the assumed isotropic synchrotron and SSC radiation fields are calculated following the results of Appendix C. Internal absorption does not play an important role in these calculations. Opacity of $\gamma$ rays from external BLR and IR photons depend on the location of the $\gamma$-ray source and characteristic size of the target photon field \citep{2012ApJ...755..147D,2012arXiv1209.2291T}. 
{ Absorption on H Ly $\alpha$ and other BLR photons can be important above $\gtrsim 25$ GeV,  and
at even lower energies on high-ionization lines such as He II Ly $\alpha$ if  $\gamma$-ray production 
takes place deep within the BLR  \citep{2010ApJ...717L.118P,sp11}.  When $\gamma\gamma$ opacity is important, 
cascade emission may contribute to the SED. 
We neglect opacity of VHE $\gamma$ rays through the BLR and IR torus in the calculations, but return to this issue
in Section 6.6.
 }

\begin{figure}[t]
% \vspace*{-2.0 cm}
\begin{center}
 \includegraphics[width=3.50in]{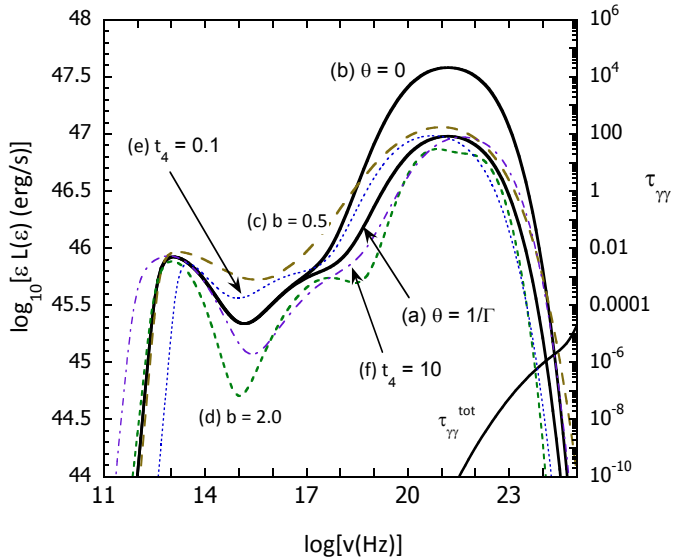} 
\vspace*{+0.2 cm}
 \caption{
Similar to Fig.\ \ref{fig1}, except that $\nu L_\nu^{pk,syn} = 10^{46}$ erg s$^{-1}$. 
Effects of varying $b$ and $t_4$ on the SED are shown as labeled. Opacity is shown
for case (a). }
\label{fig2}
\end{center}
\end{figure}

Figure \ref{fig2} shows a calculation similar to Figure \ref{fig1}, but with an order-of-magnitude smaller
apparent synchrotron luminosity, now with $\nu L_\nu^{pk,syn} = 10^{46}$ erg s$^{-1}$, but with the 
same synchrotron peak frequency. The values of 
$\Gamma$ for standard off-axis and on-axis cases (a) and (b) become smaller than in the comparable cases shown in Figure \ref{fig1}, 
but the energy densities of the external radiation fields must then become
larger in order to produce the same Compton dominance (Table \ref{table2}). The absolute jet powers become
smaller, though not decreasing as rapidly as $L_{abs}\propto L_s^{-1}$. Again, the equipartition solution ($\zeta_e = 1$)
with $t_4 = 1$ gives a breaking GeV spectrum from target Ly$\alpha$ photons, while the 
upscattered IR photons peak at  $\sim 1$ -- 10 MeV in a $\nu F_\nu$ SED,  breaking at $\gtrsim 10$ -- 100 MeV. 

For comparison, we consider how different values of $b= 0.5,1.0$, and 2.0, and 
$t_4 = 0.1$ and 10 affect the SED and values of derived quantities in the equipartition model. 
Varying $b$ changes the UV and hard X-ray fluxes by larger factors than the GeV $\gamma$-ray flux.
%compared to the Compton SED because of Klein-Nishina effects
%for $\gamma$-ray production. 
The self-absorption frequency is strongly dependent on $t_4$, which determines
the emission region size. See Tables \ref{table1} and  \ref{table2} for quantitative values and results.

\begin{figure}
% \vspace*{-2.0 cm}
\begin{center}
 \includegraphics[width=3.5in]{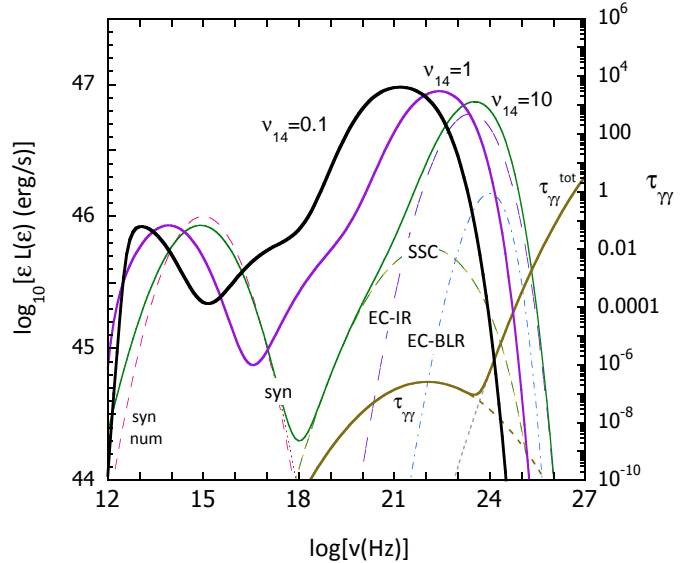} 
\vspace*{+0.2 cm}
 \caption{
Equipartition models with input parameters characteristic of ISP blazars like  3C 66A and W Comae. 
Here  $\nu L_\nu^{pk,syn} = 10^{46}$ erg s$^{-1}$, $\zeta_s = 1$, $\zeta_* = 10$ for the IR and BLR 
radiation fields, and synchrotron peak frequencies
 $\nu_{14} = 0.1, 1$, and 10  are considered. 
%Vertical lines show the range bounding peak synchrotron frequencies $\nu_s$ defining  ISP blazars, defined by $10^{14}$ Hz $< \nu_s < 10^{15}$ Hz, defined according to \citet{2010ApJ...715..429A}. 
Separate spectral 
components and the $\tau_{\gamma\gamma}$ opacity curve are shown for the $\nu_{14} = 10$ case. 
}
\label{fig3}
\end{center}
\end{figure}

Figure \ref{fig3} shows calculations for parameters characteristic of Intermediate Synchrotron Peaked (ISP) blazars, defined  \citep{2010ApJ...715..429A} as blazars with observed peak synchrotron frequencies in the range $10^{14}$ Hz $< \nu^{obs}_s < 10^{15}$ Hz.  The synchrotron and external Compton parameters are kept as before, with $\zeta_s = 1$ and $\zeta_*$ = 10 for each 
of the two external radiation fields, while varying the value of $\nu_{14}$ from 0.1 to 10. The different
SEDs for target IR and Ly$\alpha$ radiation fields are shown by the thin and thick curves for the $\nu_{14}=10$ case.
%Values of $\nu_s$ that give $10^{14}$ Hz $\lesssim \nu_s \lesssim 10^{15}$ Hz are really about an order of magnitude smaller for $b = 1$, as explained above. 
%The discrepancy decreases with increasing $b$,  and becomes exact in the limit $b \rightarrow \infty$.  
As can be seen from Tables  \ref{table1} and \ref{table2}, larger values  of $\nu_s$ imply larger values of $\Gamma$ and $\gamma_{pk}^\prime$, the combined 
effect of which greatly increases Compton-scatter power in the Thomson regime (for a fixed external field energy density), while at the same time emission at GeV energies and higher becomes
strongly inhibited  by Klein-Nishina effects when  scattering external UV  photons, so that SEDs could become increasingly dominated by $\gamma$ rays produced from target IR torus photons in the larger $\nu_{14}$ cases. 
%For the case b ISP blazar, the peak and cutoff from scattered IR photons is near 1 GeV.  If in fact, the dust torus was the dominant target photon source for the blazar $\gamma$-ray SED, then a linear correlation between the  $\gamma$-ray and synchrotron peak frequencies is predicted for blazars with $10^{13}$ Hz $\lesssim \nu_s\lesssim 10^{16}$ Hz. 
For these ISP-type parameters, structure may be found in the GeV regime due to the competition between SSC and Compton-scattered Ly$\alpha$ photons under conditions where the 
Compton dominance is not too large.  An inverted spectrum can also be formed at X-ray energies where the
$\gamma$ rays from Compton-scattered dust radiation starts to dominate the SSC emission. Such a feature may be present in the SED of 3C 66A \citep{2011ApJ...726...43A}, but will require better data and fitting to establish.

\begin{figure}
%\vspace*{+0.5 cm}
\begin{center}
 \includegraphics[width=3.5in]{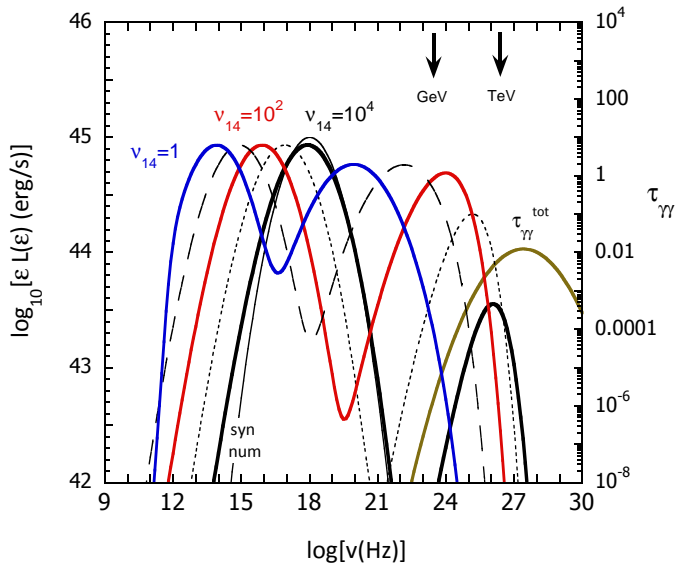} 
%\vspace*{+0.2 cm}
 \caption{
Equipartition SSC models characteristic of high-synchrotron peaked BL Lac objects.
The model shows results for  $\nu L_\nu^{pksyn} = 10^{45}$ erg s$^{-1}$, $\zeta_s = 1$, and $\log\nu_{14}=0,1,2,3,4$, 
corresponding to curves (a) through (e), respectively (see Table \ref{table1}). 
The analytic synchrotron solution, Equation (\ref{eLsyne}), is also shown for case (e). 
%Vertical lines define the range of 
%average peak synchrotron and Compton frequencies for Mrk 421 and Mrk 501. The $\nu L_\nu^{pk,syn}$ is typically between $\approx 10^{44}$ -- $10^{45}$ erg s$^{-1}$, while the peak Compton luminosity $\nu L_\nu^{pk,{\rm C}} \cong 10^{44}$ erg s$^{-1}$; see  \citet{2011ApJ...736..131A} and \citet{2011ApJ...727..129A}.  
}
\label{fig4}
\end{center}
\end{figure}

Figure \ref{fig4} shows equipartition models for blazars with characteristics similar to those of HSP BL Lac objects. Comparison of 
 models \ref{fig4}(c) -- \ref{fig4}(e) with the SEDs of Mrk 421 and Mrk 501 
from multiwavelength campaigns of  \citet{2011ApJ...736..131A} and \citet{2011ApJ...727..129A}, respectively,  suggest that 
these TeV BL Lac objects  can be well fit with our equipartition model with different values of $b$, but
quantitative statements will require a dedicated study. 
%The broader widths of the synchrotron and Compton SED components in 
%Mrk 501 compared to Mrk 421 may result from a
%smaller $b$ in the log-parabola description of the electron energy distribution, which could also 
% explain the differences between $\nu_{\rm C}$ and $\nu L_{\nu,{\rm C}}$ in
%these two sources. 

For this parameter set, we also calculated the energy density of Ly$\alpha$ radiation needed to make a significant contribution 
to the $\gamma$-ray fluxes. When $u_{Ly\alpha}\gtrsim 10^{-4}$ erg cm$^{-3}$, the emission would make an 
EC component with $\nu L_{\nu_{\rm C}} \gtrsim 10^{41.5}$ erg s$^{-1}$, just barely visible in the $\nu L_\nu$ spectrum.
The Ly$\alpha$ energy densities in Mrk 421 and Mrk 501 are, however, far below this level.
\citet{2011ApJ...732..113S} report Ly$\alpha$ luminosities of 2.4 and 5.2 $\times 10^{40}$ erg s$^{-1}$ for Mrk 421 and Mrk 501, respectively. The total jet power is $\approx 10^{44}$ erg s$^{-1}$ for the models in Fig.\ \ref{fig4}, 
and the accretion-disk luminosity $L_d$ should be greater than this if most
of the jet energy comes from mass accretion (black-hole rotational
energy could supply a comparable amount). Even for a BLR as small as 0.01 pc, 
the Ly$\alpha$ energy density $u_{Ly\alpha} \approx 10^{40}L_{40}/4\pi R^2_{0.01pc}c \approx 3\times 10^{-5}$ erg cm$^{-3}$, 
far smaller than what is necessary to make the EC $\gamma$-ray flux brighter than the SSC flux.   

The low Ly$\alpha$ luminosities of Mrk 421, Mrk 501, and also PKS 2005-489 and PKS 2155-304 \citep{2011ApJ...732..113S}, two other HSP BL Lac objects,are in accord with the thinking that TeV blazars have a very tenuous BLR environment, for example, by their formation history \citep{bd02}.
%, which may in fact disappear at sufficiently low disk luminosities \citep{2009ApJ...701L..91E}. 
By contrast, FSRQs can have very large Ly$\alpha$ luminosities. In the case of 3C 454.3, it reached 
$\approx (2$ -- 4)$\times 10^{45}$ erg s$^{-1}$  \citep{2011MNRAS.410..368B}.

%%%%%%%%%%%%%%%%%%%%%%%%%%%%%%%%%%%%%%%%%%%%%%%%%%%%%%%%%%%%%%%%%%%%%%
%%%%%%%%%%%%%%%%%%%%%%%%%%%%%%%%%%%%%%%%%%%%%%%%%%%%%%%%%%%%%%%%%%%%%%
\section{Modeling the SEDs of 3C 279} \label{sec:modelfitstodata}
%%%%%%%%%%%%%%%%%%%%%%%%%%%%%%%%%%%%%%%%%%%%%%%%%%%%%%%%%%%%%%%%%%%%%%
%%%%%%%%%%%%%%%%%%%%%%%%%%%%%%%%%%%%%%%%%%%%%%%%%%%%%%%%%%%%%%%%%%%%%%

 \citet{2012ApJ...754..114H} have organized 3C 279 campaigns around Fermi-LAT with great results. The data in Figures \ref{fig5} and \ref{fig6} show SEDs from quasi-simultaneous observing campaigns for four periods of Fermi-LAT observation, namely Epochs A (MJD 54682 - 54729; 4 Aug 2008 -- 19 Sept 2008), B (MJD 54789 - 54809; 19 Nov 2008 -- 9 Dec 2008), C (MJD 54827-54877; 27 Dec 2008 -- 15 Feb 2009), and D (MJD 54880-54885; 18 Feb 2009 -- 23 Feb 2009). 
%Similar results are obtained when fitting to later periods. 
The SEDs comprise X-ray data from Suzaku,\footnote{{\it Suzaku} consists of the XIS (0.3-12 keV), HXD/PIN (10-700 keV), and HXD/GSO (40-600 keV).} Swift XRT, XMM Newton and RXTE, optical/UV data from Kanata, GROND and Swift UVOT (170-650 nm), IR data from Spitzer, and radio data from CARMA and OVRO. For comparison,
also shown in the lower panel of Fig.\ 6 are data from the  nonsimultaneous VHE  MAGIC detection of  flare from 3C 279 flare in 2007 January \citep{2008Sci...320.1752M,2011A&A...530A...4A,2013ApJ...764..119S}. 
%The data in any case are not strictly simultaneous, as different telescopes require different observing intervals to achieve the displayed sensitivity. 
%Nevertheless, these SEDs  give some of the most detailed FSRQ spectra yet measured.  

\subsection{Accretion disk and thermal dust radiation spectrum}

A complete model of the blazar SED requires, for consistency, the emission spectrum from the accretion disk 
and from the dust. The accretion-disk spectral luminosity  is assumed to be described by a Shakura-Sunyaev disk spectrum given by
\begin{equation}
\e L_{disk}(\e) = 1.12 L_{disk}\,\left({\e\over \e_{max} }\right )^{4/3}\exp(-\e/\e_{max})\;,
\label{estarLad}
\end{equation}
normalized such that $\int_0^\infty d\e L(\e) = L_{disk}$ ($1.12 \cong 1/\Gamma(4/3) $). The $\nu F_\nu$ spectrum of the accretion-disk is therefore $f_{\e^{obs}}^{ad} = \e L_{disk}(\e)/4\pi d_L^2$, where $\e = (1+z)\e^{obs}$ and $d_L$ is the luminosity distance. The value of $\e_{max}$ depends on the spin of the black hole and relative Eddington luminosity, but for simplicity we let  $\e_{max}= 10$ eV, typical of the characteristic temperature of the UV bump in Seyfert galaxies.

The spectral luminosity of the IR dust component is approximated by a thermal distribution 
normalized to the IR luminosity $L_{IR}$. Thus 
\begin{equation}
\e L_{IR}(\e) = {15 L_{IR}\over \pi^4}\; {(\e/\Theta)^4\over \exp(\e/\Theta) - 1}\;,
\label{estarLdust}
\end{equation}
with corresponding $\nu F_\nu$ spectrum  $f_{\e^{obs}}^{IR} = \e L_{IR}(\e)/4\pi d_L^2$.
Letting $\Theta = k_{\rm B} T/m_e c^2 = 2\times 10^{-7}$ corresponds to $T = 1200$ K,
but note carefully that the mean photon energy used in the monochromatic approximation
is a factor $\Gamma(4)\zeta(4)/\Gamma(3)\zeta(3) \cong 2.70$ 
larger than $\Theta$. The IR luminosity spectrum and components for the warm dust model
shown in Figs.\ 5 and 6 correspond to a dust temperature
$T\cong 440$ K, a factor $2.7$ less than 1200 K, so that 
$2.70\times \Theta = 2\times 10^{-7}$, as well as the 1200 K dust.

A dust covering factor of 20\% is assumed, so that
$L_{IR} = 0.2 L_{disk}$. The energy density of the IR radiation is limited by the energy density of a blackbody, namely 
$$u_{bb}({\rm erg~cm}^{-3})(T) = 9.36\times 10^{24}\Theta^4 $$
\begin{equation}
\cong 3\times 10^{-4} ({T\over 440{\rm~K}})^4  \cong 0.016 ({T\over 1200{\rm~K}})^4\;.
\label{ubb}
\end{equation}
Thus the energy density at most reaches $\approx 10^{-3}$ erg cm$^{-3}$ for warm dust, 
and $\approx 0.02$ erg cm$^{-3}$ for hot dust.

\begin{figure}
\vspace*{+0.5 cm}
\begin{center}
 \includegraphics[width=3.5in]{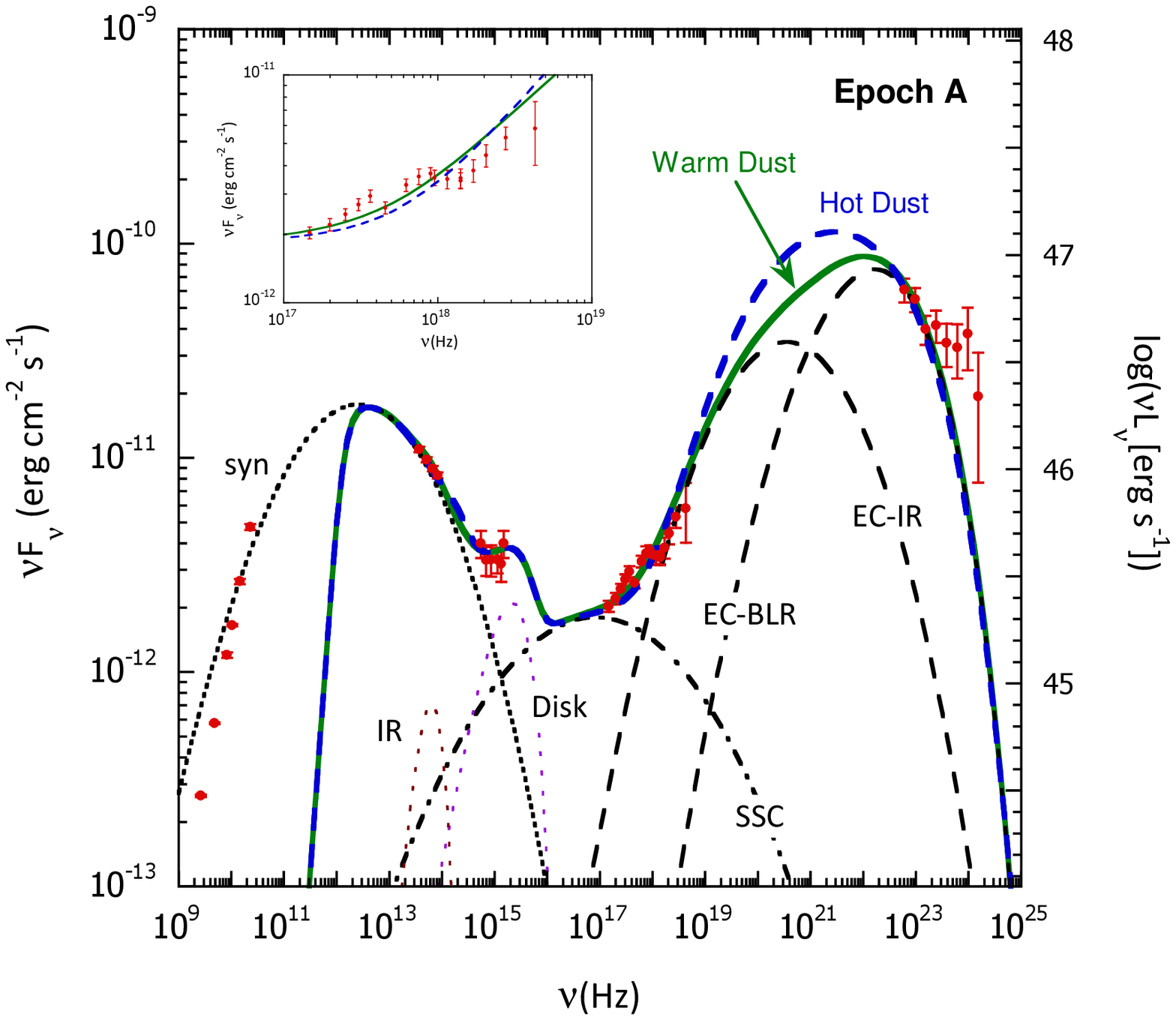} 
 \includegraphics[width=3.5in]{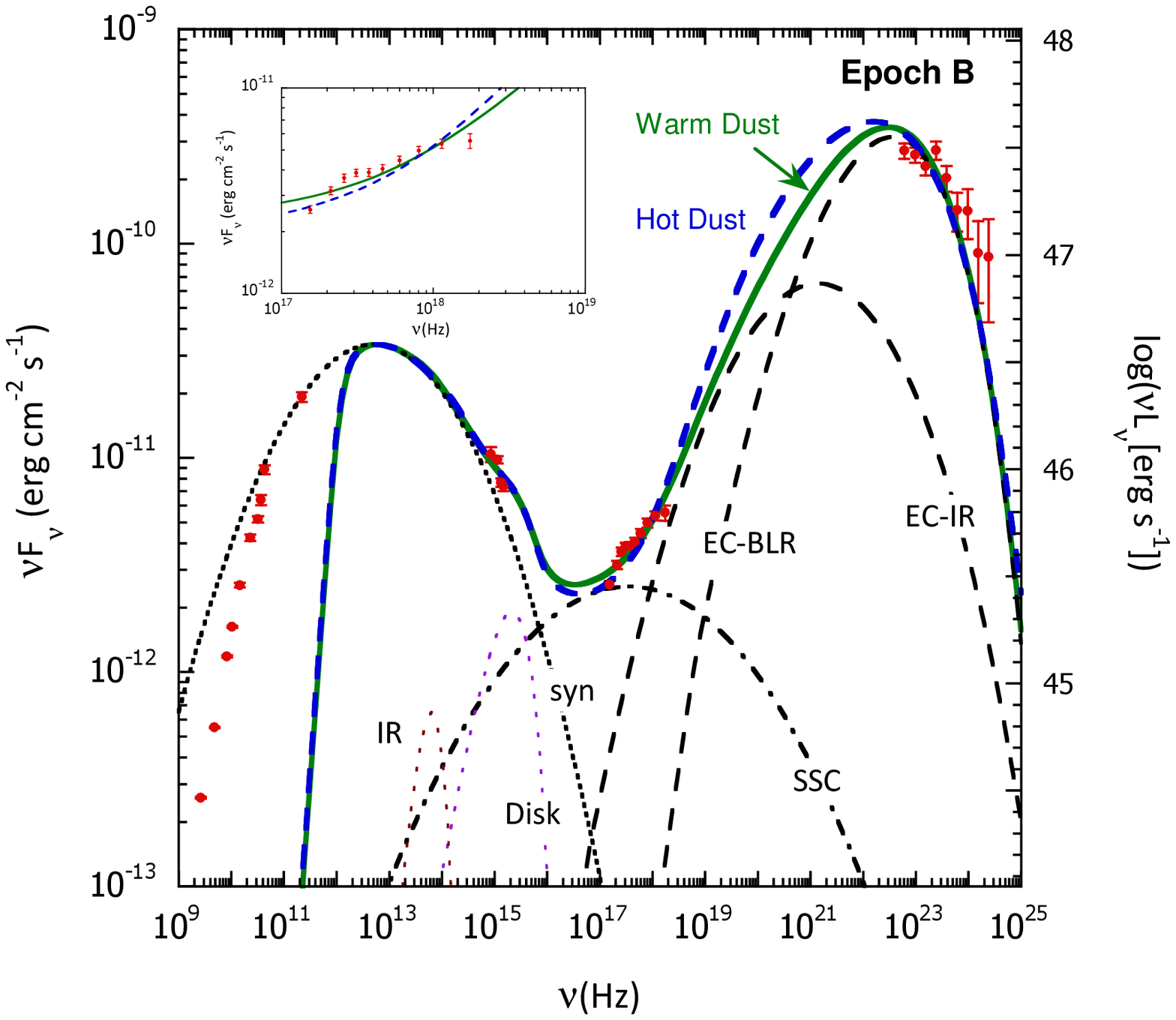} 
%\vspace*{+0.2 cm}
 \caption{
Equipartition blazar model fits to the SEDs of 3C 279 \citep{2012ApJ...754..114H}
for Epochs A (upper panel) and B (lower panel), 
with input parameters given in Table \ref{table3} and implied properties from the
model in Table \ref{table4}.  Two spectral fits corresponding to warm-dust and hot-dust 
IR radiation fields are considered. Separate components are shown, with the EC-IR and 
EC-BLR components shown for the warm-dust solution.  Insets show detail of fits at X-ray energies.
}
\label{fig5}
\end{center}
\end{figure}

\begin{figure}
\vspace*{+0.5 cm}
\begin{center}
 \includegraphics[width=3.5in]{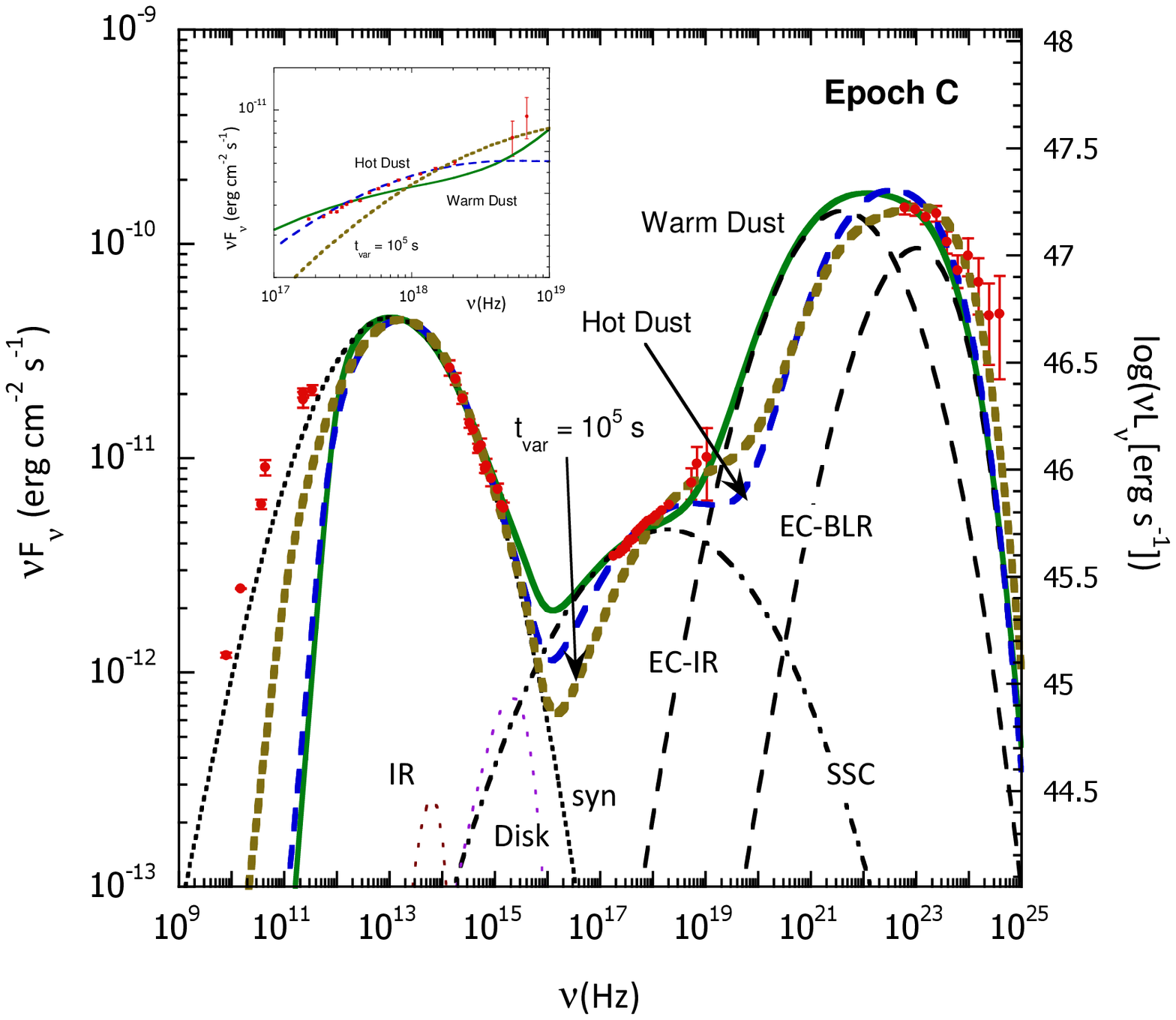} 
 \includegraphics[width=3.5in]{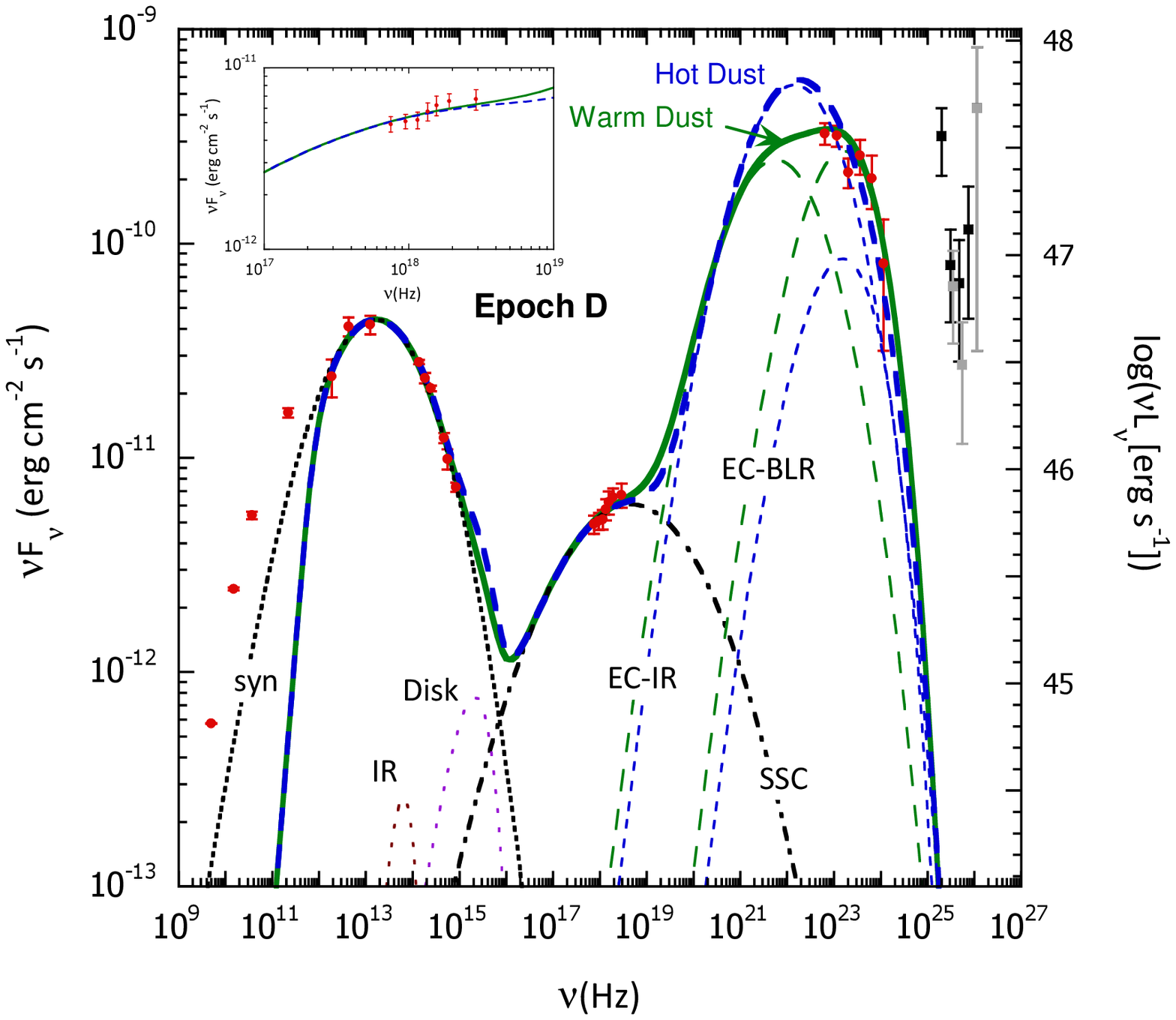} 
%\vspace*{+0.2 cm}
 \caption{
Equipartition blazar model fits to the SEDs of 3C 279 \citep{2012ApJ...754..114H}
for Epochs C (upper panel) and D (lower panel), with spectral components and 
inset graphs as described in Fig.\ \ref{fig5}. The upper panel shows a warm dust, hot dust,
and long variability time, $t_{var} = 10^5$ s, solution. The lower panel for the Epoch 
D fit shows both  warm-dust and hot-dust Compton scattering components. 
Non-simultaneous VHE MAGIC data for 3C 279
are shown in the lower panel for comparison with the Epoch D 
{\it Fermi}-LAT $\gamma$-ray data. 
}
\label{fig6}
\end{center}
\end{figure}

\begin{table}[t]
\begin{center}
\caption{Blazar input parameters for 3C 279$^a$}
\begin{tabular}{llllllll}
\hline
%\vskip0.1in
% &   &   & n  &  & $\nu$ &    & $\nu_\beta$ &  \\
Epoch & $\nu L_\nu^{pk,syn}$ & $\nu_{14}$ & $b$ & $\zeta_s$ & $\zeta_{IR}$ &  $\zeta_{BLR}$ & $L_{disk}$ \\ 
 & ($10^{48}$ &  &  &  &  & & ($10^{46}$   \\ 
 & erg s$^{-1}$) &  &  &  &  & & erg s$^{-1}$)  \\ 
\hline
\hline
% &  &  &   &   &  &   \\ 
Aw$^b$ & 0.023 & $0.05$ & $0.8$   & 0.15 & 2.5 & 7 & 0.55  \\
% &  &  &  &  & &   & &   \\ 
Ah$^c$ & &  &   &   & 6 & 5 &   \\
Bw & 0.043 & 0.1 & 0.6 &  0.11 & 2.5 & 18& 0.5 \\
% &  &  &   &  &  &   & &  \\ 
Bh& &  &  & 0.1 & 8 & 12 & 0.55 \\
% &  &  &   &  &  &   & &  \\ 
% &  &  &  &  & &   & &   \\ 
Cw & 0.06  & 0.2 & 1 & 0.15 & 4 & 5 & 0.2  \\
Ch &    & 0.3 & 1.3 & 0.2 & & &   \\
Ct5$^a$ &    & &  & 0.28 & 8& 3&   \\
% &  &  &   &  & &   & &  \\ 
Dw &   & &  & & 7 & 16 &  \\
Dh &   & &  &  & 16 & 5 & 0.5 \\
% &  &  &   &  & &    \\ 
\hline\end{tabular}
\label{table3}
\end{center}
$^a$$t_{4}= \zeta_e = \zeta_{p,nuc} = N_\Gamma = 1$, except for Ct5, where $t_4 = 10$\\
$^b$w: Warm dust solution: $\epsilon_{BLR} = 2\times 10^{-5}$, $T\cong 440$K \\
$^c$h: Hot dust solution: $\epsilon_{BLR} = 5.4\times 10^{-5}$, $T \cong 1200$ K\\
\end{table}

\begin{table}[t]
\begin{center}
\caption{Model output values for 3C 279}
\begin{tabular}{ccccccc}
\hline
% &   &   & n  &  & $\nu$ &    & $\nu_\beta$ &  \\
Epoch & $\Gamma, \dD$ & $B^\prime$  & $\gamma^\prime_{pk}$ &  $u_{IR}$ & 
 $u_{{\rm Ly}\alpha }$&  
$\log[L_{B,par}$ \\ 
 &   &    (G) &  & ($10^{-3}{\rm~erg}$ &  ($10^{-3}{\rm~erg}$  & $(L_{ph})$  \\
 &   &      &  & ${\rm cm}^{-3})/$ & ${\rm cm}^{-3})$ &(erg s$^{-1}$)]\\
 &   &      &  &  $T_{min}({\rm K})^a$ & &\\
\hline
\hline
% &  &  &   &   &  &   \\ 
%1 & 8.9 & 17.8 & 6.7 & 140 & 0.172 & $5.33$ & 45.0 \\
Aw & 25.7 &  2.6 & 31.6 & $0.77/560 $ & $2.16$& $45.8(45.8)$  \\
Ah &   &    &   & 1.85/690 & 1.5  & $45.8(45.9)$  \\
Bw& 35.0 & 1.8  & 28.7 & $0.19/400$& $1.4$ &   $46.0(46.1)$  \\
Bh & 36.5 & 1.7 & 29 & $0.49/500$ & $0.74$ &   $46.0(46.1)$  \\
Cw & 37.2 & 1.3 & 99 & 0.15/370 &  0.19 &   45.8(45.8)  \\
Ch & 34.8 &  & 164 & 0.17/390 &  0.21 &   45.7(45.8)  \\
Ct5 & 22.5 & 0.41 & 364 & 0.03/250 &  0.08 &   45.9(46.1)  \\
Dw &  &  &  & 0.30/440   & 0.68&   $45.7(46.1)$  \\
Dh &  & &  & 0.68/550  & 0.21&   $45.7(46.2)$  \\
% &  &  &   &   &  &  \\ 
\hline\end{tabular}
\label{table4}
\end{center}
$^a$Minimum blackbody temperature to exceed $u_{IR}$\\
\\
\end{table}

\subsection{Fits to Epochs A-D of 3C 279}

Figures \ref{fig5} and \ref{fig6}  show fits to 3C 279 from our equipartition model. 
The values of $\nu L_\nu^{pk,syn}$ and $\nu_{s}$ are input, and
$\zeta_s$ and the external radiation field parameters are varied in order
to fit the data. 
The assumed value $t_{var}= 10^4$ s then constrains the system to  
 give  $\delta_{\rm D},$ $B^\prime$ and $r_b^\prime$
through the equipartition equations, 
Equations (\ref{deltaD}), (\ref{BprimeG}), and (\ref{gammapprime}), respectively.

With the goal of limiting the number of parameters, we take $t_4 = \zeta_e = \zeta_{p,nuc} = N_\Gamma = 1$, and 
vary $\zeta_s $, $\zeta_{BLR}$ and $\zeta_{IR}$, which normalize the synchrotron, external BLR and IR radiation-field  
energy densities, respectively, until a reasonable fit to the 
broadband data is achieved. 
In fact, $t_{var}$ can be measured, as can $N_\Gamma$ if the angle  to the jet axis 
can be determined, and such values should be used when available.  
The value $\zeta_{p,nuc}$ only affects the jet power, not the spectrum,
as is proved from an examination of Equations (\ref{u'tot}), (\ref{Ljet}), and (\ref{deltaD}) -- (\ref{gammapprime}). 
We keep $\zeta_e$ equal to unity in this study, this condition taken to define equipartition.
%up parameter space widely, deserves further study, but will be deferred to later study. 

In the fitting, there is freedom to assign the accretion-disk temperature and luminosity, 
and therefore the IR dust luminosity for the  assumed 20\% covering
factor. Moreover, the accretion-disk emission can be important for fitting structure
in the optical wavelength range, but in other cases gives only an upper limit to disk luminosity $L_{disk}$. 

The results of fitting the 3C 279 data for Epochs A-D are shown in Figures \ref{fig5} and \ref{fig6} and 
listed in Tables \ref{table3} and \ref{table4}. The spectral components and total spectra are shown in these figures for the  case
of a warm dust IR field ($\epsilon_{IR} = 2\times 10^{-7}$). A second fit with a hot-dust
($\epsilon_{IR} = 5.4\times 10^{-7}$) photon field is also shown for each of the four
epochs (heavy dashed curves). The different components, as labeled, are the 
unabsorbed and self-absorbed nonthermal synchrotron spectra, the IR dust field,
 the Shakura-Sunyaev accretion-disk field (big blue bump), 
the SSC radiation, EC $\gamma$ rays from jet electrons scattering either IR photons radiated by 
dust (EC-IR) or Ly$\alpha$ radiation (EC-BLR).
Except at the highest $\gamma$-ray energies, 
reasonable fits with either warm and hot dust seem to be found.

Given the equipartition assumption, spectral fitting implies the energy densities of the 
external radiation field, given in Table \ref{table4}. There we also see the minimum 
blackbody temperature that makes a radiation field with energy density as implied by 
the spectral fits. We conclude that the warm-dust models are narrowly allowed, when
one considers that if only 20\% of the accretion-disk radiation is re-radiated as dust, 
then the dust temperature has to be a factor $\sim 5^{1/4} \sim 1.5$ larger than calculated
in Table \ref{table4}. The covering factor for warm dust could also be 
somewhat larger than $>20$\%.

The significance of the implied energy densities and jet parameters
are discussed in the next section. As just remarked, the fit is not satisfactory at the highest 
{\it Fermi}-LAT energies. Indeed, as shown in the bottom panel of Fig.\ \ref{fig6},
the equipartition model has no hope of fitting the MAGIC VHE data. We discuss this 
further below. The insets show the quality of the spectral-model fits to the X-ray 
data. There is a suggestion of a component peaking at $\sim 5$ keV, but whether
this could be related to the SSC component or hot plasma surrounding the supermassive black 
holes is unclear. 

%%%%%%%%%%%%%%%%%%%%%%%%%%%%%%%%%%%%%%%%%%%%%%%%%%%%%%%%%%%%%%%%%%%%%%
%%%%%%%%%%%%%%%%%%%%%%%%%%%%%%%%%%%%%%%%%%%%%%%%%%%%%%%%%%%%%%%%%%%%%%
\section{Discussion} \label{sec:discussion}
%%%%%%%%%%%%%%%%%%%%%%%%%%%%%%%%%%%%%%%%%%%%%%%%%%%%%%%%%%%%%%%%%%%%%%
%%%%%%%%%%%%%%%%%%%%%%%%%%%%%%%%%%%%%%%%%%%%%%%%%%%%%%%%%%%%%%%%%%%%%%

{ By equipartition we mean that in the fluid frame of the jet, the 
nonthermal electron and positron energy density is 
equal to the magnetic-field energy density. A system in equipartition
is close to the minimum power solution, and for that reason important to study.
Ideally, by choosing a suitable functional form for the electron spectrum,
and measuring variability time $t_{var}$, 
the number of free parameters is sufficiently constrained that fits to data
determine model parameters, including the energy densities of the 
surrounding radiation fields for external Compton scattering.
%The success of this approach depends importantly on the quality of the
%fits to the SEDs.
}

\citet{2012ApJ...754..114H}, 
analyzing multiwavelength data of 3C 279, 
find a number of interesting results relevant here.
%by this equipartition jet model. 
For example, 
a delay by about 10 days in the optical emission compared to the 
$\gamma$ rays is found from cross-correlation studies. 
Complex structure is observed at near- and far-IR frequencies, and
an inverted spectrum is seen near $3\times 10^{12}$ Hz 
from the radio/IR data in Epoch D, with a smoothly curving synchrotron spectrum through 
the synchrotron $\nu F_\nu$ peak. The extension of the synchrotron spectrum from 
radio/IR into  optical/UV frequencies 
 indicates a hardening in Epochs A and possibly B, as well as Epochs F and H.

Quasi-thermal radiations from the accretion disk
could make the excess optical emission 
and apparent synchrotron hardenings.
In the modeling, Shakura-Sunyaev accretion-disk 
emissions were used to fit the optical spectra in  
Epochs A and B (Figure 5). This nuclear blue-bump radiation must 
exist in order to illuminate the BLR \citep{1994ApJ...421..153S},
and provides a powerful photon source for Compton scattering
\citep{1993ApJ...416..458D} by nonthermal
electrons in the inner jet, within the BLR.\footnote{The transition radius from the
near field to far field, beyond which the 
accretion disk can be treated as a point source, is at $r_{NF\rightarrow FF}\cong 0.5
M_9(\Gamma/10)^4$ pc, and the transition from the dominance of 
accretion-disk  to scattered radiation is at
$r_{NF\rightarrow sc} \cong 0.01 M_9^{1/3} [r_{BLR}/({\rm 0.1~pc})]^{2/3}/(\tau/0.1)^{1/3}$ pc 
 \citep{2002ApJ...575..667D}. Except in the inner jet, or for blazars lacking a BLR,
 the dominant
target radiation field is the BLR or, farther out, the IR torus field
\citep{2009ApJ...704...38S}.   }
Joint IUE/ASCA observations
 of 3C 279  during a low $\gamma$-ray state
\citep{pia99} found excess optical/UV radiation consistent with an accretion-disk
spectrum, with disk luminosity $L_d\approx 2\times 10^{45}$ erg s$^{-1}$ and temperature reaching 20,000 K. We use 
$L_d\approx (2$ -- $5)\times 10^{45}$ erg s$^{-1}$ in the equipartition model for
3C 279 (Table 4).

Because of synchrotron self-absorption, 
radio emission longward of 300$\mu$ ($\approx 10^{12}$ Hz) is strongly self-absorbed in the 
one-zone model used here to explain blazar radiations.
Long-wavelength radio emission therefore must originate
 from a larger volume surrounding the jet. 
A change in 
spectral and variability properties at sub-mm and longer wavelengths
would be expected at the transition from optically thick to optically thin regimes.
An interesting study is the dependence of the
SSA frequency on model parameters, following
Equation (\ref{kappaepsilon1}).

%: the nonthermal
%synchrotron component for the nonthermal radio/mm/IR/optical radiations; 
%accretion-disk radiations to explain the optical/UV blue bump; nonthermal SSC X-rays;
%hard X-rays observed by the Swift XRT and BAT and  NuSTAR, made by SSC and Compton-scattered
%IR (EC-IR) photons originating from outside the jet; 
%and GeV $\gamma$ rays by external Compton scattering of Ly$\alpha$ and BLR photons (EC-BLR). 
%If the models lacked the EC-IR $\gamma$-ray component, the 
%model SEDs would be very different, both at the lower end of the Fermi-LAT range and 
%in the hard X-ray/soft $\gamma$-ray spectrum explored by Swift BAT, Suzaku, INTEGRAL, and NuSTAR. 
%Note that accretion-disk
%radiations are required to obtain good fits to the optical spectra in Epochs A and B,
%and the model does not apply to the $\lesssim 10^{12}$ Hz radio measurements. 
%The available model constraints match the number of
%model parameters, except for the fr.

%\subsection{Model Implications}

%In this approach to multiwavelength FSRQ modeling, fitting to a blazar SED implies  values of $\delta_{\rm D}$%, $B^\prime$, 
%$\gamma_{pk}^\prime$, and $r_b^\prime$ for a specific value of $t_{var}$ and observing angle $\theta$. 
Our fits
to the multi-epoch data of 3C 279, found in Figs. \ref{fig5} and \ref{fig6} and Tables \ref{table3} and \ref{table4}, imply  $\Gamma\approx \delta_{\rm D} \approx 25$ -- 37
and $B^\prime \approx$ few G, 
assuming $\theta = 1/\Gamma$ and $t_4 = 1$. These values can be compared with parameters used in modeling of 
earlier 3C 279 data by 
\citet{2000ApJ...545..107B}\footnote{$B^\prime = 0.81$ G, $\Gamma =$ 20, $\theta = 1/\Gamma = 2.86^\circ$, $\gamma_{br} = 150$, with $t_{var}\approx 20$ d.}
and  
\citet{2001ApJ...553..683H}\footnote{$B^\prime = 1.5$ G, $\Gamma = 5$ -- 15, corresponding to $t_{var} \approx 2$ -- 6 d variability timescales; observing angle $\theta = 2^\circ$ from \citet{1999ApJ...521..493L}.} 
when fitting to multiwavelength data associated with earlier EGRET campaigns, where
reasonable fits were obtained by  
 choosing values of $\Gamma$ and $B^\prime$. The parameters we deduce are, however, quite different from the values used to fit 
the same data sets in the paper by  \citet{2012ApJ...754..114H}. For 
an observing angle of $\sim 2^\circ$
and a 3 -- 4$^\circ$ opening angle jet \citep{2004AJ....127.3115J,2005AJ....130.1418J},
they chosse $B^\prime \cong 0.15$ G and $\Gamma\approx $ 15 -- 20 to fit the data, which
requires large characteristic electron Lorentz factors that allows IR photons to be scattered to GeV energies. 
%One of the reasons is the assumed observing angle with respect to the jet direction.
The main difference is the long assumed variability time scale in the modeling
of  \citet{2012ApJ...754..114H}, namely $t_{var} \sim 15$ d. 
The trend to such long timescales appears consistent with what we find in Fig. 6, Epoch C, 
when we relax the variability time to $t_5 = 1$ (i.e., $t_{var} \cong $1 d in run Ct5), 
so that longer timescales of variability make scattering to higher $\gamma$-ray energies more feasible. 
The relevant question then is, 
what value of $t_{var}$ should be 
used in the various epochs?

\subsection{Location of $\gamma$-ray production site and variability}

For only a few  blazars  does the {\it Fermi} LAT have sensitivity to probe to a few hour time
scale during major outbursts, namely  3C 454.3 \citep{2011ApJ...733L..26A}, 
PKS 1222+216 \citep{2011ApJ...733...19T},
PKS 1510-089 \citep{2013ApJ...766L..11S,2013MNRAS.431..824B},
and 3C 273 \citep[see also][]{2011A&A...530A..77F,2013MNRAS.430.1324N}. Detection 
of much more rapid variability, as short as $\approx 5$ min in PKS 2155-304 \citep{2007ApJ...664L..71A}, 
and a few minutes in in Mrk 501 \citep{2007ApJ...669..862A} and Mrk 421 \citep{2012AIPC.1505..514F}, 
could be peculiar to the TeV BL Lac objects,
%represents a major puzzle, which we'll return to later.
%perhaps involving magnetic reconnection. 
but 4C +21.35 has varied at 70 GeV -- 400 GeV energies on timescales
as short as 30 minutes. By comparison, significant MAGIC VHE detections of 3C 279 took place on two successive 
days, 22 and 23 February 2006 \citep{2008Sci...320.1752M}, 
after which it was also detected during  a flare on 16 Jan 2007 \citep{2011A&A...530A...4A}. 
The VHE fluxes for the two days in 2006 
were each significant, and 
 differed by $>2\sigma$, indicating day-scale variability.
%VHE variability shorter than a day was not demonstrated, nor have there been 
%any record of recent VHE detections of 3C 279.}
 
The observing epochs we examine 
have durations of  6 days (Epoch D), 3 weeks (Epoch B), and 6 weeks (Epochs A and C). 
%and statistics accumulated during these observing times are reflected in the size of
%the error bars in the Fermi-LAT SEDs
%shown in Figs.\ \ref{fig5} and \ref{fig6}  \citep{2012ApJ...754..114H}.
The derived energy densities of the target radiation fields for Epochs A -- D of 3C 279 are 
$u_{IR}\approx (0.1$ -- 1) $\times 10^{-3}$ erg cm$^{-3}$ 
for the IR field and
$u_{{\rm Ly}\alpha}\approx (0.2$ -- 2)$\times 10^{-3}$ erg cm$^{-3}$ for the Ly$\alpha$ field, except
for the third case in Epoch C where $t_{var} = 10^5$ s and the implied external radiation-field 
energy densities
are much lower. Close inspection of the light curves from radio through
$\gamma$ rays in
\cite{2012ApJ...754..114H} indicates variability on a timescale as short as a fraction of a day to a few days, making 
the $t_4 \cong 1$ -- 10 regime most relevant for spectral fitting. 
A better fit to the $\gamma$-ray data and a worse fit to the X-ray data
is found with a longer variability time. 
     
%The Ly$\alpha$ field comprises $\approx 1/3$ of the total broad line flux \citep{2002ApJ...565..773T},
%The other strong lines are generally close enough to the Ly$\alpha$ frequency that the monochromatic 
%approximation remains a good approximation to the composite scattered spectrum \citep{cer13}.

%The large, $\Gamma \gtrsim 20$ bulk Lorentz factors found in the equipartition model for 3C 279 are consistent with 
%it being transparent to GeV $\gamma$ rays, so that opacity and cascade emission can be neglected. On the other hand, 
%hadronic secondary $\gamma$ radiations  can be severely absorbed by the IR dust field \citep{2012ApJ...755..147D}. 
%A further check on the model is to make sure that $\gamma$ rays can escape the emission region by 
%avoiding internal $\gamma\gamma$ absorption with the synchrotron radiation, and external $\gamma\gamma$ absorption with the Ly$\alpha$ and BLR radiation.

The total external-field 
energy densities
vary by about an order of magnitude, from  $\approx (0.4$ -- 3) $\times 10^{-3}$ erg cm$^{-3}$, 
for $t_{var} \approx 10^4$ s. We can use our results to define minimum radii
for $\gamma$-ray production, using the expression
\begin{equation}
u_{BLR} = {\tau L_{disk}\over 4\pi r^2_{BLR} c [1+(r/r_{BLR})^{\beta_{BLR}}]}\simeq 
{0.3\tau\over (1+\rho^{3})}\,{{\rm erg}\over {\rm cm}^3}\;
\label{uBLR}
\end{equation}
\citep{2009ApJ...704...38S,2012ApJ...754..114H}
and assuming the simple scaling relation 
 \citep{2008MNRAS.387.1669G}
\begin{equation}
r_{\rm BLR} \cong 0.1 \sqrt{ {L_{\rm disk}\over 10^{46}{\rm ~erg~s}^{-1}}}\;{\rm pc}\;
\label{RBLR}  
\end{equation}
between the characteristic BLR radius, $r_{\rm BLR}$, and
 accretion-disk luminosity  $L_{\rm disk}$. If a fraction $\tau\approx 0.1$ of the radiation is scattered or reprocessed into line radiation, 
 we find that the $\gamma$-ray emission site takes place at $\rho = r/r_{BLR} \sim 2$ -- 10, 
 that is, $\sim 0.1$ -- 0.5 pc from the black hole.

%\subsection{Jet Model}

Within the colliding-shell paradigm for blazars \citep[e.g.,][]{spa01,bd10,ma12}, the collision radius $R_{coll} \lesssim 2\Gamma^2 c t_{var}\sim 0.2 (\Gamma/30)^2 t_4$ pc, which is consistent with 
a location at the edge of the BLR (Eq.\ (\ref{RBLR})). 
In Epoch A, when the derived $\Gamma$ is smallest (see Table \ref{table2}),
so $R_{coll} \lesssim  0.1 (\Gamma/26)^2 t_4$ pc, the probable location of the emission 
region is deepest within the BLR and IR torus, and the implied energy densities are largest at this epoch. 
Given the uncertainty on $t_4$, the colliding-shell radius is also on the $\sim 0.1$ -- 0.5 pc size scale 
of the IR dust emission where 
$u_{IR}$ might change significantly, depending on the geometry of the IR torus.
The scatter in IR energy density also reflects the range of model uncertainty.\footnote{The data 
gap in the 0.1 -- 100 MeV range reflects sensitivity limitations of MeV telescopes.  INTEGRAL 
observations of 3C 279 were made by \citet{2010A&A...522A..66C}, but unfortunately
at the time when SPI was being annealed.}

From \citet{2008ApJ...685..147N,2008ApJ...685..160N}, 
the inner radius due to dust sublimation is 
\begin{equation}
R_d \simeq 0.4 \sqrt{L_{45}} ({T_{sub}\over 1500 {\rm~K}})^{-2.6}\;{\rm~pc}\;,
\label{Rdsublimation}
\end{equation}
where the bolometric AGN luminosity is $10^{45}$ erg s$^{-1}$, 
and $T_{sub}$ is the dust sublimation temperature.
 \citet{2011ApJ...732..116M} argue for dust sublimation radii of 1 -- 2 pc. 
The IR zone might not be much larger than the BLR, however, as would
 naturally follow if the torus and BLR clouds are the same accretion flow found at different
distances from the nuclear continuum, with properties defined by whether the 
accreting material was inside or outside the dust-sublimation radius
 \citep{2008ApJ...685..160N}.

%Here the energy density of the radiation fields is also greatest, which takes place if the collision radius were smallest, 
%and most within the radiation zone. 

%The largest value of $u_{IR}\approx 2\times 10^{-3}$ erg cm$^{-3}$ was deduced in Epoch A, 
%which also had smallest derived value of $\Gamma$. 
%to argue that the characteristic size of the IR torus could be as small as $\approx 0.4$ pc. 
% The $\gamma$-ray emission at later Epochs could be made at  $\gtrsim 0.5$ pc from the black hole, where the %IR field is smaller, consistent with the %
%smaller energy densities of radiations inferred in later Epochs.

\subsection{Jet power calculations}

The absolute two-sided jet power $L_{jet}$(erg s$^{-1}$) is given by Equation (\ref{Ljet}),
%$L_{jet} = 2\pi r_b^{\prime 2} \beta \Gamma^2  c u_{tot}^\prime$,
%with $u_{tot}^\prime = u^\prime_{B^\prime} (1+ \zeta_e  +\zeta_{p,nuc} + \zeta_{s}+\zeta_*)$ for the total 
%energy density, from  Equation (\ref{u'tot}).  The jet power $L_{jet}$(erg s$^{-1}$) is 
in terms of magnetic-field/particle power $L_{B,par}$  and photon power $L_{ph}$, as 
reported in Table \ref{table4} for the four epochs of 3C 279, assuming $\zeta_{p,nuc} = 1$. 
%We also give the absolute photon
%powers $L_{ph} = L_{abs,syn}+L_{abs,C}$, from Equations (\ref{Ljetsyn}) and (\ref{LjetEC}) for the synchrotron/SSC
%and EC photon powers, respectively. 
%A value of $\zeta_{p/nuc} = 1$ is 
%used in the values of $L_{jet}$ given in Table \ref{table4}. 
%The maximum
%jet power found in our calculations
%, again with baryon-loading unity, 
%is $L_{jet} \approx 10^{46.3}$ erg s$^{-1} \approx 2\times 10^{46}$ erg s$^{-1}$, which 
%should be compared with the Eddington luminosity.
Emission-line studies using 
H$\beta$ line width  \citep{2001MNRAS.327.1111G} and broad line luminosity
 \citep{2002ApJ...579..530W}
implies that 3C 279 harbors a black hole with mass (3 -- 8)$\times 10^8 M_\odot$.
The implied Eddington luminosity is therefore in the range $(4$ -- 10)$\times 10^{46}$
erg s$^{-1}$. 
%If Eddington conditions limit maximum
%jet power, given that
%then the most powerful flare is at the Eddington limit (provided 
%$\zeta_{p/nuc}$ is sufficiently small). 
All our solutions  have total powers  below $\approx 2\times 10^{46}$ erg s$^{-1}$, 
well below the Eddington luminosity, but a factor of a few higher than the accretion-disk
luminosity.
%(which could have a broadened emission spectrum when modified blackbody 
%or advective effects are considered).
%(fit Bw has a total power of $\approx 10^{47}$ erg s$^{-1}$), 
%and often closer to $10^{46}$ erg s$^{-1}$.
%The absolute jet powers for the warm-dust solutions, in units of $10^{47}$ erg s$^{-1}$, 
%the maximum Eddington luminosity, are $\zeta_{p/nuc}=(1, 10, 100)=$
%$(0.14,0.2,0.95)$ for Epoch A, $(0.54,0.71,2.5)$ for Epoch B, 
%$(0.26,0.46,2.4)$ for Epoch C, and $(0.46,0.63,2.3)$ for Epoch D.
%To keep the absolute jet luminosity 
%$\lesssim 10^{47}$ erg s$^{-1}$ in all cases requires a b
We find that baryon-loading factors
$\zeta_{p,nuc}\lesssim 20$ are compatible with an Eddington-limited system.
  The photon power is about equal to the 
combined magnetic-field and particle power in the jets of 3C 279; see Table \ref{table4}.
%This is contrary to the blazars in our model study, where the photon power dominates.
%WHY?

The apparent isotropic bolometric  $\gamma$-ray luminosity exceeds $\approx 10^{48}$ erg s$^{-1}$ in Epoch B, whereas
the total jet power is $\lesssim 2\times 10^{46}$ erg s$^{-1}$ (for  $\zeta_{p,nuc} = 1$).   
If the jet opening angle $\theta \approx 1/\Gamma$, then the difference between
the absolute and apparent powers would be proportional to the 
beaming factor $f_b$, with $f_b \lesssim $0.1\% for $ \Gamma \approx 30$.
The reduction in power is by only 2\% rather than 0.1\%, which shows that in relativistic nonthermal synchrotron
jet models, it can be naive to use a simple beaming factor to relate the apparent isotropic 
and absolute powers.

%The reduction in power is by only 2\% rather than 0.1\%, because the equipartition condition is applied to the synchrotron component, 
%which has a bolometric luminosity of $\approx 10^{47}$ erg s$^{-1}$, an order of magnitude
%less luminous than the $\gamma$ rays. 
%The total jet power
%is found to be a factor of a few greater than the accretion-disk luminosity.
%which is puzzling

\subsection{Leptonic models for $\gamma$-ray emission from 3C 279}

The equipartition leptonic model, as presented in Figures 5 and 6, simulates 
reasonably well the measured SEDs of 3C 279, though the model underproduces
multi-GeV $\gamma$-ray 
fluxes. 
 A better BLR model including the line structure does not change the shape
enough to explain the discrepancy \citep{cer13}.
It is interesting
to consider how the discrepancy can be corrected in a leptonic 
model. 
One possibility is a hardening in the electron spectrum compared to 
the monotonic softening in the log-parabola
function. This could
help explain the optical/UV hardening at the same time as enhancing  fluxes at 
GeV energies, and could work for Epochs A and B, but
not for Epoch C, where there is no evidence for  hardening at optical/UV energies.
Increasing the variability time as in run Ct5 allows the leptonic model to 
fit the $\gamma$ rays, but the X-rays now require an additional component
to compensate at soft X-ray energies. Corrections to the X-ray spectrum 
could involve complicating factors, e.g., 
absorption, or addition of emission from a hot plasma surrounding the nucleus.

This leptonic model is therefore not entirely satisfactory, and 
faces another challenge when trying to explain VHE emission detected
from 3C 279. Modeling by \citet{2009ApJ...703.1168B} showed
that leptonic models for the MAGIC data required leptonic models
well out equipartition that give bad fits to the X-rays.
Nevertheless, leptonic models have considerable success in fitting the 
broadband SED of 3C 279 in most states, so the basic blazar model
may only need extended.

In the modeling performed in the  
\citet{2012ApJ...754..114H} paper,  a conventional
blazar model is used, much as described here except with a doubly 
broken power law to describe the electron distribution. 
By using a long, $\approx 2$ week variability time, 
a smaller fluid magnetic field, $B^\prime \approx 0.15$ G,
and smaller bulk Lorentz factors, $\Gamma\approx 15$ -- 20, can be
used, though the characteristic
electron Lorentz factors must be larger.
This permits IR photons to be scattered to higher energies before the
Klein-Nishina suppression sets in.
%This model fits well the optical and $\gamma$-ray data.
Except for Epoch A, where a broken power-law model is used and 
a good fit to the entire data set is achieved, the other models in   
\citet{2012ApJ...754..114H}
do not account for the X-ray data because, as argued there, the X-rays do not correlate
with the $\gamma$-ray and optical fluxes during flaring states.
%If the X-rays are primarily SSC, a correlated variability behavior, 
%different for SSC and EC $\gamma$ rays, can be expected.
Their short variability model D2 requires an unexpectedly large energy 
density ($\approx 10$ erg cm$^{-3}$), so this model can be 
ruled out.

In the model by \citet{2012MNRAS.419.1660S}, first-year Fermi data 
of 3C 279 \citep{2010ApJ...716...30A} is fit 
with $\Gamma=26,B^\prime = 0.67$ G$, u_{IR} = 4\times 10^{-3}$ erg cm$^{-3}$, an 860 K dust temperature,
and a one-day variability time scale.
Scattered accretion-disk radiation gives a similar 
emission component \citep{2001ApJ...553..683H,2002ApJ...575..667D} as IR
radiation, so it is important to determine if the $\gamma$-ray emission 
region is made deep within the BLR or farther out, where the BLR radiation is
more dilute. Here $\gamma\gamma$ studies at the VHE regime can determine 
location.

%An important goal of the Cherenkov Telescope Array is to 
%to search for rapid variability from 3C 279 to match the 
%variabiity time and size scale of the system 
%with the emission frequency, and determine whether this is 
% allowed in a given leptonic model. 

\subsection{Particle acceleration and the highest energy $\gamma$ rays}

The log-parabola function,  { originally introduced
because it gave good fits to the curving synchrotron SEDs of ``active" (blazar-like) radio sources
\citep{1986ApJ...308...78L} and to the $\gamma$-ray spectra
of Mrk 421 and Mrk 501 \citep{1999ApJ...511..149K}, has since been used to fit 
blazar \citep{2002babs.conf...63G,2004A&A...413..489M} 
and gamma-ray burst \citep{2010ApJ...714L.299M} spectra.}   It has also been recognized as 
a convenient functional form for the electron energy distribution \citep{2006A&A...448..861M}.
%\citet{2011ApJ...739...66T} examine the properties of this function in detail, and  show how this form can arise in a second-order acceleration scenario.

If a log-parabola nonthermal electron spectrum is capable of 
fitting multiwavelength blazar data reasonably well, as we 
have tried to demonstrate here, then one can ask
about the physical processes that can generate 
such a distribution. Curving functions arise
in second-order, stochastic acceleration scenarios 
\citep[e.g.,][]{1995ApJ...446..699P,2006ApJ...647..539B,2011ApJ...739...66T}. 
%If the quality of fits are better 
%and if fewer parameters are used in the current framework than 
%employed in shock/injection scenarios, we may
%conclude that the curving spectrum is a better model for 
%the data.
%A shock framework may still be required in order
%to provide a mechanism for generating turbulence.
A turbulent scenario 
%producing curving particle spectra 
might also naturally arise in
 models involving relativistic shocks and jet reconnection in 
Poynting-dominated jet models and compact 
knots far ($\gtrsim$ pc) from the nucleus \citep[e.g.,][]{2009MNRAS.395L..29G,nal12}. 
{ The accelerated particle distribution is, however, poorly known, and dependent on 
model assumptions.} 
%So this approach to fitting blazar data might
%also be related to current reconnection models.
% An interesting question is why the
%$b=1$ solution is a good first approximation to the electron
%spectrum.

With regard to the fits to 3C 279, the equipartition 
model underproduces the $\gg$ GeV
emission measured by {\it Fermi}-LAT in 
Epochs A, B, and C, but can be corrected with a long variability time, 
at the expense of fitting the X-ray data.
The VHE detection of 3C 279 \citep{2008Sci...320.1752M} would, however,
have to  require either emissions from leptons far out of equilibrium
accompanied by poor fits to the X-ray or synchrotron 
data.
%\footnote{TeV radiation can be made an equipartition model; see
%Fig.\ \ref{fig4}, but would be accompanied by a 
%synchrotron radiation component peaking at X-ray energies; 
%cf.\ the Mrk 501 flare \citep{1998ApJ...492L..17P}. }
Another possibility is hadronic processes
to explain the $\gamma$-ray excess.

\subsection{Hadronic $\gamma$ rays in 3C 279 and UHECRs}

In view of emission 
extending to the VHE range, which is hard to understand
with leptonic models \citep{2009ApJ...703.1168B}, a new spectral component
may be required. A possible
candidate is  UHECRs accelerated in blazar jets. 
As we have seen, the jet powers can accommodate large
baryon-load factors, $\zeta_{p,nuc}\lesssim 20$, so 3C 279 or indeed 
other powerful FSRQs, could display
high-energy hadronic emission tails \citep{2003ApJ...586...79A,2009ApJ...703.1168B,2013ApJ...768...54B}}.

From the \citet{1984ARA&A..22..425H} condition, 
the maximum particle energy 
is limited to energy  
$ E < E_{max}\cong Z c e \delta_{\rm D}^2 B^\prime t_{var}$.
Using Equations (\ref{deltaD}) and (\ref{BprimeG}), we find
\begin{equation}
E_{max}({\rm eV})\cong 
1.4\times 10^{20} Z\; {L_{48}^{5/16} t_{4}^{1/8} f_1^{1/4} f_2^{1/8}\over \zeta_e^{1/4} \zeta_s^{1/6} f_0^{1/16}}\;,
\label{Emax}
\end{equation}
so equipartition blazars with sufficiently large apparent power can accelerate
protons to ultra-high energies  \citep{2010ApJ...724.1366D,2009ApJ...690L..14M}. Thus 3C 279 
could in principle accelerate protons to super-GZK energies by this basic requirement. FSRQs like 3C 279 are, however, unlikely to make all the UHECRs, because FSRQs are not found within the GZK radius of $\sim 100$ -- 200 Mpc. Moreover, in spite of their great power, the low space density of FSRQs and FR2 radio galaxies makes them unlikely to be the primary source class powering the UHECRs, compared to 
 FR1 radio galaxies and BL Lac objects, which however have trouble to accelerate  $\gtrsim 10^{19}$ eV protons \citep{2012ApJ...749...63M}. 

UHECR production in FSRQs like 3C 279 might be revealed by detection of
steady, extended cascade radiation induced by photopion
% production from escaping
%UHECR protons. At redshift $z = 0.5362$, the angular size distance
%$d_A = 1301.8$ Mpc ($H_0 = 71$ km (s-Mpc)$^{-1}$, 
%$\Omega_m = 0.27 = 1-\Omega_\Lambda$). Depositing energy on a 
%distance scale of 100 -- 200 Mpc with a jetted beam pointed 
%nearly along our line of sight will create through photopion and 
and photopair processes from beamed ultra-relativistic protons
travelling through intergalactic space, as proposed to explain the spectra 
and variability properties of
some unusual TeV BL Lac objects
%a steady, extended $\gamma$-ray emission feature 
\citep{2010APh....33...81E,2010PhRvL.104n1102E,2012ApJ...751L..11E}.
Provided the
 UHECR beam can escape from the structured regions surrounding
3C 279 without being dispersed \citep{2012ApJ...745L..16M}, 
a slowly varying UHECR-induced $\gamma$-ray halo should surround 3C 279. The MAGIC
detection of  VHE emission from 3C 279 shows, however, VHE emission
that possibly varies on timescales less than a day.
Hadronic models considered by \citet{2009ApJ...703.1168B} to fit the MAGIC VHE 
data require large powers. 
It remains to be seen if modifications of leptonic 
scenarios, for example, involving strongly magnetized jets 
and magnetic reconnection models in a multi-zone jets-within-jets--type model \citep{2010arXiv1005.5551M},
or hadronic models with proton synchrotron or photopion
production and cascades \citep{2013ApJ...768...54B}, are preferred to make the VHE $\gamma$-ray spectra in 3C 279. 

%A blazar like 3C 279, in the ecliptic plane at latitude 0.2$^\circ$, 
%is not optimally located for IceCube observations, but otherwise would 
%be a promising target for PeV neutrino searches. The external Ly$\alpha$ radiation
%field provides a target for ultra-relativistic protons to make a neutrino
%spectrum down to PeV energies (see Fig.\ 3 in \citep{2012ApJ...755..147D};      
%contrary to statements of \cite{2012arXiv1211.1974C}). %

%Hadronic secondary $\gamma$ radiations  can be severely absorbed
%by the IR dust field \citep{2012ApJ...755..147D}.

%1996A&AS..120C.503G

\subsection{External $\gamma\gamma$ Opacity and CTA}

{ Expressions for $\gamma\gamma\rightarrow$ e$^\pm$ opacity \citep{gs67,bmg73} applied to the 
inner-jet environment can be found in \citet{2012ApJ...755..147D}.
For a monochromatic isotropic external radiation field,
the $\gamma\gamma$ opacity $\tau_{\gamma\gamma}(\e_1)$ reaches its maximum value
 $\tau^{max}_{\gamma\gamma,}(\e_1^{max}) \cong 0.56\times (3\sigma_{\rm T} R/8m_ec^2\epsilon_0$) at
$\epsilon_1 \cong 3.54/\epsilon_0$. For Ly $\alpha$ photons,  the $\gamma\gamma$
opacity from an external radiation field therefore reaches its maximum for 90 GeV photons at the source, and
the condition $\tau_{\gamma\gamma} < 1$ implies an attenuation length $\lambda_{\gamma\gamma}$(pc)$ < 0.04/u_{-3}$, where the BLR energy 
density is $10^{-3}u_{-3}$ erg cm$^{-3}$.  Detection of $\approx 100$ GeV photons immediately puts
the location of the $\gamma$-ray emission site far beyond $r_{BLR}$, depending on the value of $\tau$ in Equation (\ref{uBLR}).
Consequently the VHE $\gamma$-ray production sites in 3C 279 and 4C+21.35 are far outside the BLRs of these
sources. External $\gamma\gamma$ absorption effects should however be unimportant in the Fermi 
LAT spectrum of 3C 279, measured below $\approx 10$ GeV.

The $\gamma\gamma$ opacity constraint is even more severe for attenuation by infrared dust photons, though 
for higher energy $\gamma$ rays.   For a graybody radiation field with temperature $m_ec^2\Theta/k_{\rm B}$ and 
energy density $=10^{-4} u_{-4}$ erg cm$^{-3}$,
the opacity $\tau^{IR}_{\gamma\gamma}$ reaches a maximum value of $1.076\times (45/8\pi^4) (\sigma_{\rm T}R u_0/ m_ec^2\Theta$) 
at $\epsilon \cong 2/\Theta$. For 1200 K dust, the attenuation length $\lambda_{\gamma\gamma}$(pc) $\cong 0.01/u_{-4}$ 
for $\approx 5$ TeV $\gamma$ rays. Observations of VHE emissions in blazars with ground-based $\gamma$-ray telescope arrays
and with the Cherenkov Telescope Array both at $\approx 100$ GeV and multi-TeV 
energies will be important for confining the location of the $\gamma$-ray emission site.
   }

\subsection{Blazar Correlations and Equipartition}

Besides fitting individual spectra of a single blazar,
 the principle of equipartition holds promise for explaining blazar
correlations such as the blazar sequence
and blazar divide. The blazar sequence  \citep{1998MNRAS.299..433F,1998MNRAS.301..451G} 
refers to an inverse correlation of $\gamma_{pk}^\prime$ and $L_\gamma$, though 
later studies argue for additional complexity related to 
jet structure \citep{2011ApJ...740...98M} that can be explained
in cooling scenarios \citep{2013ApJ...763..134F}.
The blazar divide refers to the correlation between the $\gamma$-ray spectral index and 
 $\gamma$-ray luminosity \citep{2010ApJ...715..429A,2009MNRAS.396L.105G}, and
is closely related to the blazar sequence, as is the correlation of $\gamma$-ray spectral 
index with synchrotron $\nu F_\nu$ peak { frequency \citep{2010ApJ...715..429A}.  The equipartition 
approach can predict trends and correlations 
in the statistics of blazars with, e.g., $t_{var}$, $L_{syn}$, and Compton dominance, 
that can be compared with data, and is currently under investigation. }

\section{Summary}

With the goal of minimizing the number of free parameters, 
we use a three-parameter log-parabola function for the electron 
distribution and assume equipartition between the energy densities of
the magnetic field and nonthermal leptons ($\zeta_e = 1$).
On this basis, we derive Equations (\ref{deltaD}) -- (\ref{gammapprime}) 
giving $\dD$, $B^\prime$, $\gamma_{pk}^\prime$. The main inputs 
from the data are $\nu L_\nu^{pk,syn}$, the peak value
of the $\nu L_\nu$ synchrotron spectrum, $\nu_s$, the frequency at which the $\nu L_\nu$ synchrotron
spectrum peaks, and $t_{var}$, the variability time scale. Using the equipartition 
assumption, values of $\zeta_s$ and $\zeta_*$, which determine 
the SSC and external Compton-scattered $\gamma$-ray fluxes, are adjusted to 
fit the multiwavelength data.

We considered 
warm dust ($T = 440$ K) and hot dust ($T = 1200$ K) IR radiation fields,
and assumed $\theta = 1/\Gamma$ and $t_{var} = 10^{4}$ s. 
From fits to 4 epochs of 3C 279,
  Ly$\alpha$ radiation energy densities 
$\sim (0.2-2)\times 10^{-3}$ erg cm$^{-3}$ are derived, and IR 
energy densities a factor of a few smaller. 
These values are compatible with a $\gamma$-ray production 
site at the outer edge of the 
 BLR, $\gtrsim 0.1$ pc from the nucleus. If the variability time
is longer than $10^4$ s, the external field energy densities can 
be smaller and distance from the nucleus larger. The absolute
jet powers are well below the Eddington limit unless the baryon-loading
is $\gtrsim 10$, and somewhat larger
than the accretion-disk luminosities used in the fits.

%Even in this approach, the simple blazar model has 10 parameters:
% (1) $\delta_{\rm D}$, (2) $\Gamma$, (3) $B^\prime$, (4) ${\cal E}^\prime_e$, 
%(5) $\gamma_{pk}^\prime$, (6) $\epsilon_{BLR}$, (7) $u_{BLR}$, (8) $\epsilon_{IR}$, 
% (9) $u_{IR}$, and (10) $b$,  which is simplified to 6 parameters by assuming 
%$\theta = 1/\Gamma$, $b = 1$, $\epsilon_{BLR}=\epsilon_{Ly\alpha}=2\times 10^{-5}$,
%and  $\epsilon_{IR}$  corresponds to either 440 K dust (warm dust, $\epsilon_{IR} = 2\times 10^{-7}$) or 
%1200 K dust (hot dust, $\epsilon_{IR} = 5.4\times 10^{-7}$). The six parameters are determined 
%by the observables (1) $\nu_s$, (2) $\nu L_{\nu_s}$, (3)  $t_{var}$, (4) $\nu_{SSC}$, (5) $\nu %L_{\nu_{\rm SSC}}$, 
%(6) $\nu_{C}$, and (7) $\nu L_{\nu_{\rm C}}$; values for parameters (4) -- (7) depend
%on modeling the overlapping components in the X-ray and $\gamma$-ray spectra.
%The best-fit model implies the energy densities of the target radiation fields. 

%is instead outside the BLR radius 
%$R_{BLR} \sim 0.1$ -- 0.2 pc. 
The inferred IR energy densities, 
ranging by about an order of magnitude for the four different epochs
considered, are easier to explain if the $\gamma$-ray emission site
is found in the same radial range where IR gradients take place. 
%over the  to explain if the IR radiation field has to vary 
%by such a large factor over the $<7$ month
% period considered here. It  therefore seems more likely 
%that the emission regions occur outside the BLR, at  
% $R\sim 0.1$ --  $0.5$ pc and  
%as we derive
% if IR gradients take place on the same radial scale. 
A location of the $\gamma$-ray emission site
$\approx 0.1$ --  $0.5$ pc from the nucleus 
is also consistent with relativistic shell collision 
radii.

We also find that protons satisfy the Hillas condition for  
acceleration above $10^{20}$ eV 
for values of $\delta_{\rm D}$,  $B^\prime$, and $r_b^\prime$ derived 
in this model for the jets of 3C 279. Whether leptonic or hadronic processes
make the $\gg$ GeV radiation observed with {\it Fermi}-LAT and 
VHE radiation observed with MAGIC is not clear. 
%Though FSRQs are not likely to explain the 
%origin of all UHECRs, UHECRs in FSRQs can make a VHE $\gamma$-ray 
%emission component such as has been detected in sources like 3C 279, 
%PKS 1222+216, and possibly PKS 1510-089. Such a feature
%would be difficult to explain in the framework of the leptonic equipartition model presented here, 
%but the uncertainty related to the variabilty timescale and equipartition
%relation makes it hard to draw definite conclusions.

To summarize, this paper introduces a new blazar modeling technique 
based on equipartition between leptons and magnetic field, assuming
the log-parabola form for the $\gamma^{\prime 2} N^\prime(\gamma^\prime)$ electron
spectrum. We tested the method by fitting recent Fermi-LAT and multiwavelength data
of 3C 279, deriving reasonable values for the broad-line and IR energy densities if 
the emission site is at $\sim 0.1$ -- 0.5 pc from the black hole, and the variability
time is $10^4$ s. 
%The modeling does not answer the question of VHE $\gamma$-ray emission,
%, and the puzzle
%of rapid variability if $\gamma$-ray emission is made far from the black hole.
Excess GeV radiation detected with the Fermi-LAT 
can be explained for larger variability times, 
but then the X-rays are not well fit.
 A sensitive MeV telescope would be valuable to determine the frequency of
the peak  and the contributions of scattered
IR emission to the $\gamma$-ray spectrum of blazars.
%A straightforward extension of the current modeling shown in the paper
% is to take $\theta $ directly inferred for a given source.
%A hard excess above the extrapolation of the Fermi-LAT 
%spectrum would, for a leptonic model, not be expected, and further
%constrained by the synchrotron SED. 
Searches for rapid variability
of VHE $\gamma$ rays from 3C 279 with imaging 
atmospheric Cherenkov gamma-ray telescopes, 
including the  Cherenkov Telescope Array, will 
be crucial for finding the 
limits of leptonic models and potential
signatures of hadrons.
%origin of the high-energy $\gamma$ rays in
%blazars. 
%In this way, the nonthermal leptonic model can be tested, and 
%if found deficient, provide an opening to test the theory of UHECR origin in blazars.  
%Observations with CTA will be useful in this regard.

%(J2000.0) 194.0465271  -5.7893119 

%Search for rapid variability
%of VHE $\gamma$ rays from 3C 279 with  Northern Hemisphere imaging 
%atmospheric Cherenkov gamma-ray 
%telescopes 

\acknowledgements 

We thank Dr.\ M.\ Hayashida for providing the 3C 279 data, and J.\ D.\ Finke and 
S.\ Razzaque for discussions. We thank the referee for an excellent and challenging report.
%The paper is better for it.
The work of C.D.D.\ is supported by the Office of Naval Research and the NASA {\it Fermi} Guest Investigator Program.
%\pagebreak

%{ 
\appendix

\section{Corrections to equipartition relations}
% with Log-Parabola Representation of the Electron Distribution}

For an electron distribution with the log-parabola form given by 
$\g^{\prime 2} N_e^\prime(\gp)= K x^{-b\log x}$ from Equation (\ref{g2Ng}),
with  $K = \gamma_{pk}^{\prime 2}N^\prime_e(\g^\prime_{pk})$ and $x =\gamma^\prime/\gamma^\prime_{pk}$, 
we can derive a few elementary relations. First, 
\begin{equation}
N_{e0} = \int_1^\infty d\gp N_e(\gp) = {K\over \gamma^\prime_{pk}}(\int_0^\infty dx \,x^{\hat b \ln x} - \int^\infty_{\gamma^\prime_{pk}} dx \,x^{\hat b \ln x}) 
\equiv {K\over \gamma^\prime_{pk}}(I - I_{e}) \;,
\label{Ne0}
\end{equation}
defining $\hat b = b/\ln 10$. 
Interestingly,
\begin{equation}
I =\int_0^\infty dx \,x^{-2-\hat b \ln x} = \int_0^\infty dx \,x^{-\hat b \ln x} = 10^{1/4b}\sqrt{\pi \ln 10\over b}
%\equiv{ 1\over 2 f_3 }\;,\;
%f_3 = {10^{-1/4b}\over 2\sqrt{\pi \ln 10/b}}.
\label{I}
\end{equation}
Thus
\begin{equation}
K =  {N_{e0} \gamma^\prime_{pk} \over I}
= {N_{e0} \gamma^\prime_{pk} 10^{-1/4b}\over \,\sqrt{\pi \ln 10/b}}\;\equiv \; 2 N_{e0}\gamma^\prime_{pk} f_3 \;. 
\label{Ne0K}
\end{equation}
The fractional error incurred by letting $\g^\prime_{pk}\rightarrow 0$ in Equation (\ref{Ne0}) is
\begin{equation}
{I_{e}\over I}  = {1\over 2}\,[1-{\rm erf}(u)]\;\asympt\; {\exp(-u^2)\over 2u\sqrt{\pi}}\;,
%\;{\buildrel \rightarrow \over {_{\hat b \gg 1}}}\; \sqrt{b\over\pi \ln 10}\,10^{-1/4b}\;.
\label{IeI}
\end{equation}
where erf(u)$\equiv {2\over \sqrt{\pi}}\int_0^z dt \exp(-t^2)$ is the 
 error function, and $u = \sqrt{\hat b} \ln \gamma_{pk}^\prime + 1/2\sqrt{\hat b}$.
The error is always small when $\gamma^\prime_{pk} \gg \exp{\sqrt{1/\hat b}}$ which,
for typical values $b\approx 1$ obtained in the fits, means that the correction is negligible
when $\gamma^\prime_{pk} \gg 5$.
%The final expression is obtained by taking the limit that $\hat b, b \ll 1$.

The total nonthermal electron energy in the comoving fluid frame is 
\begin{equation}
{\cal E}^\prime_{e} = m_ec^2\int_1^\infty d\gp \gp N_e(\gp) \;
={K m_ec^2}(\int_{1/\gamma_{pk}^\prime}^\infty dy \,y^{-1-\hat b \ln y} - \int_{\gamma^\prime_{pk}}^\infty dy \,y^{-1-\hat b \ln x}) 
\equiv K( I_1 - I_{1e})\;. 
\label{EeprimeK}
\end{equation}
It follows that
\begin{equation}
I_1 = \sqrt{\pi \ln 10\over b}\;\;.
%\;\;{I_{1e}\over I_1} \;,
%[1-{1\over 2} \sqrt{b\over \ln 10} \,{\rm erfc}(\ln \gamma_{pk}^\prime )]\cong K m_ec^2 \sqrt{\pi \ln 10\over b}\;=\;  N_{e0} f_1 m_ec^2 \gamma_{pk}^\prime\;,
\label{Ee0}
\end{equation}
The fractional error $I_{1e}/I_1$ incurred by letting $\gamma_{pk}^\prime \rightarrow 0$ is given by  the right hand side of 
 Equation (\ref{IeI}), except that now $u = \sqrt{\hat b } \ln \gamma_{pk}^\prime $. When $\gamma^\prime_{pk} \gg \exp(1/\sqrt{\hat b })$,
this error is negligible. Comparison with values of $b\approx 1$ and $\gamma_{pk}^\prime\gtrsim 20$ for the considered cases
shows that neglecting the error correction is a good approximation.

%with $f_1\equiv 10^{-1/4b}$. The error is of order $\sqrt{b/\ln 10}\, (\gamma_{pk}^{\prime -\ln\gamma_{pk}^\prime}/ \ln \gamma_{pk}^\prime) $, which is always small when  $\gamma_{pk}^\prime \gg 1$ and $b \gg 0.01$. The Lorentz factor corresponding to the mean electron energy is $\langle \gamma_e^\prime \rangle = {\cal E}^\prime_e/m_ec^2 N_{e0} =\gamma_{pk}^\prime/ 10^{1/4b}$.

\section{$\delta$-function synchrotron spectrum for log-parabola electron distribution}
 
Here we go beyond the mono-energetic electron approximations for the 
equipartition relations, and include corrections for the width parameter $b$
of the log-parabola function.

Beginning with  the log-parabola electron distribution, 
%$\g^{\prime 2} N_e^\prime(\gp)= K x^{-b\log x}$ 
Equation (\ref{g2Ng}),
%Beginning with electron distribution, Equation (\ref{g2Ng1}), and 
and the comoving synchrotron luminosity from mono-energetic electrons, 
 $L^\prime_{syn} = {4\over 3} c\sigma_{\rm T} (B^{\prime 2}/8\pi) \gamma^{\prime 2}N_{e0}$, it is straightforward to derive  
\begin{equation}
\epsilon L_{syn}(\epsilon ) = \upsilon  x^{1-\hat b \ln x}\;
\label{eLsyne}
\end{equation} 
for  the spectral synchrotron luminosity. 
To obtain this result, we use 
the $\delta$-function approximation for the mean synchrotron photon energy, 
$\epsilon_{syn}^\prime = (3/2) (B^\prime/B_{cr})\gamma_{pk}^{\prime 2}$, and the relation 
$\epsilon^\prime L_{syn}(\epsilon^\prime ) = \gamma^\prime L^\prime_{syn}(\gamma^\prime )$. 
Here  $x\equiv \sqrt{\epsilon/ \epsilon_{pk}}=\sqrt{\epsilon^\prime/ \epsilon^\prime_{pk}}$, 
and $\epsilon_{pk}= \delta_{D} \epsilon_{pk}^\prime=(3/2)\delta_{D} (B^\prime/B_{cr}) \gamma_{pk}^{\prime 2}$
is the peak synchrotron photon energy from a mono-energetic electron distribution with Lorentz factor $\gamma^\prime_{pk}$.
The upsilon coefficient
\begin{equation}
\upsilon \equiv K\,\delta_{\rm D}^4 c \sigma_{\rm T} {B^{\prime 2}\over 12\pi} \gamma_{pk}^\prime  
=\, L_{syn}f_3 
%\;=\;10^{1/4b}f_3 \, {{\cal E}_e^\prime\over m_ec^2 }\;L_{syn}
\;,\;
\label{upsilon}
\end{equation}
using  Equations (\ref{Ne0K}) and (\ref{Ee0}), assuming negligible error. 

The dimensionless peak frequency $\epsilon_s$ of the $\nu F_\nu$ synchrotron spectrum is obtained by taking the maximum of Equation (\ref{eLsyne}).
One finds 
\begin{equation}
\epsilon_s = f_2\epsilon_{pk}\;,\; f_2\equiv  10^{1/b}\;. 
\label{epsilons}
\end{equation}

The bolometric synchrotron luminosity, neglecting self-absorption corrections, is given by
Equation (\ref{eLsyne}) through
\begin{equation}
L_{syn} = \int_0^\infty d\epsilon\, L_{syn}(\epsilon )  = {4\over 3}\,c\sigma_{\rm T}N_{e0} \delta_{\rm D}^4 \,u_{B^\prime}^\prime \gamma_{pk}^{\prime 2}\; = \;
{\upsilon\over f_3} \;.
\label{Lsyncorr}
\end{equation}
Equation (\ref{Ls}) for the synchrotron luminosity is recovered without correction.
The relationship between $L_{syn}$ and the $\nu L_\nu$ peak synchrotron luminosity
at $\epsilon = \epsilon_s$ is 
\begin{equation}
L_{syn} = 2 \sqrt{{\pi \ln 10\over b}} \, \nu L_{\nu}^{pk,syn}\;, 
\label{Lsyn}
\end{equation}
so, from eq.\ (\ref{upsilon}),
\begin{equation}
\upsilon \equiv 10^{-1/4b}\,\nu L_{\nu}^{pk,syn}\;.
\label{upsilon1}
\end{equation}

Starting from Equation (6.60) in \citet{dm09} and using the 
$\delta$-function Thomson approximation (Equation (6.58), which is exact in the Thomson limit,
one can derive the ratio of the apparent SSC and synchrotron luminosities. 
We find
\begin{equation}
{L_{SSC}\over L_{syn}} = {\sigma_{\rm T} N_{e0}\gamma_{pk}^{\prime 2}\over 3\pi r_b^{\prime 2} f_0} 
= {\sigma_{\rm T} {\cal E}^\prime_{e}\gamma_{pk}^\prime\over 3\pi m_ec^4 \delta_{\rm D}^2 t_{var}^2 f_0 f_1 }=
{4\over 3} {c\sigma_{\rm T} u^\prime_e \gamma_{pk}^\prime\over m_ec^2 f_1} \;.
\label{LsscoverLsyn}
\end{equation} 
%, using The final term on the right hand side expresses 
%coefficient $\upsilon$ in terms of ${\cal E}_e^\prime$, using Equation (\ref{Ee0}).
%$\nu F_\nu$ dimensionless peak frequency of the synchrotron component.

\section{$\gamma\gamma$ Opacity of Synchrotron Radiation Emitted by log-parabola electron distribution}

Using Equation (\ref{eLsyne}) and the relation $\epsilon L_{syn}(\epsilon ) =  \delta_{\rm D}^4 \epsilon^\prime L_{syn}^\prime (\epsilon^\prime )$, the comoving spectral photon density
\begin{equation}
n^\prime (\ep ) = {\epsilon^\prime u^\prime (\epsilon^\prime )\over m_ec^2 \epsilon^{\prime 2}} = {\upsilon x^{1-\hat b \ln x}\over 
4\pi f_0 m_ec^3 r_b^{\prime 2} \epsilon^{\prime 2} \delta_{\rm D}^4}\;.
\label{nprimeep}
\end{equation}
The $\gamma\gamma$ opacity for a photon with comoving energy $\epsilon_1^\prime = \epsilon_1/\delta_{\rm D}$ traveling through an emission region with characteristic size $r_b^\prime$ is
\begin{equation}
\tau_{\gamma\gamma}(\epsilon_1^\prime ) \cong r_b^\prime \int_0^\infty d\ep\,n^\prime(\ep ) \sigma_{\gamma\gamma}(\ep \epsilon_1^\prime)\;
= {\upsilon \sigma_{\rm T} \epsilon_1^\prime\over 24\pi f_0 m_ec^4 t_{var} \delta_{\rm D}^5}\, \hat x^{1-\hat b \ln \hat x }\;\equiv\;  
 {\cal C} \hat x^{1-\hat b \ln \hat x }\;,
\label{taugg}
\end{equation}
using the $\delta$-function approximation, $\sigma_{\gamma\gamma}(\ep\epsilon_1^\prime) \cong 2\sigma_{\rm T} \delta(\epsilon^\prime \epsilon_1^\prime -2)/3$,
for the $\gamma\gamma$ cross section. 
Here $\hat x = \sqrt{\ep/\epsilon^\prime_{pk}} = \sqrt{2/\epsilon_1^\prime \epsilon_{pk}^\prime}= \delta_{\rm D} \sqrt{2/\epsilon_{1}\epsilon_{pk}}$. The peak $\gamma\gamma$ opacity occurs at $\hat x_{max} = 10^{-1/2b}$, or at photon energy 
$\epsilon_{1,max} = 2\delta_{\rm D}^2 10^{2/b}/\epsilon_{s} $, as can be found by differentiating Equation (\ref{taugg}) with respect to $\epsilon_1$. Thus $\tau_{\gamma\gamma}(\epsilon_{1} ) 
= \tau_{\gamma\gamma,max} \hat x^{-3-\hat b \ln \hat x }$, where $\tau_{\gamma\gamma,max} = 10^{-1/4b}{\cal C}$, and
\begin{equation}
{\cal C} =  {10^{-1/4b}\sigma_{\rm T}  \epsilon_1 (\nu L_\nu^{pk,syn}) \over 24 \pi \,m_ec^4 f_0 t_{var} \delta_{\rm D}^6 }
\cong 1200\epsilon_1\;{10^{-1/4b}(\nu L_\nu^{pk,syn}/10^{48}{\rm ~erg~s}^{-1}) \over 
f_0 t_{4}\delta_{\rm D}^6}\;,
\label{tggmax}
\end{equation}
using Equation (\ref{epsilons}). When $\epsilon_1 \gg 1$, the peak opacity can be large, depending on $\delta_{\rm D}$, but only at very large photon energies, namely
\begin{equation}
E_{1,max} = m_ec^2 \epsilon_{1,max} \cong  {10^{17}{\rm~eV}\over \nu_{14}} \; \left( {\delta_{\rm D}\over 30}\right)^2\, \left ({10^{2/b}\over 10^2}\right)\;.
\label{E1max}
\end{equation}

For the $\gamma\gamma$ opacity of a photon in the SSC radiation field, Equation (\ref{taugg}) is used with $\epsilon_s$ replaced by the peak SSC photon energy $\epsilon_{SSC}$
where the $\nu L_\nu$ SSC luminosity reaches its maximum value at $\nu L_\nu^{pk,SSC}$, 
and  $\nu L_\nu^{pk,syn}$ replaced by $\nu L_\nu^{pk,SSC}$. We assume that the log-parabola width parameter for the 
SSC SED spectrum is $\approx b/2$.
%These results are used to plot the $\gamma\gamma$ opacities in the figures.

%}

%The  apparent isotropic bolometric synchrotron 
%({\g^\prime\over\g^\prime_{pk}})^{-b\log({\g^\prime\over \gamma_{pk}^\prime})}
%\clearpage
\bibliographystyle{apj}
\bibliography{blazar}

%\bibliography{apj-jour,myrefs}

\end{document}